\documentclass{aa}  

\usepackage{graphicx}
\usepackage{txfonts}
\usepackage{lipsum}
\usepackage{subcaption}   
\usepackage{lscape}   
\usepackage{placeins}     

\begin{document}

   \title{Solar photospheric spectrum microvariability }

   \subtitle{III. Radial velocities and line profiles in magnetic active-region granulation}
  
   \author{Dainis Dravins
          \inst{1}, 
       Hans-G\"{u}nter Ludwig
           \inst{2},
           Matthias Steffen
           \inst{3},
           Carlos Allende Prieto
           \inst{4}, and
       Lars Koesterke
           \inst{5}
           }
%
% Please retain full first names of authors! 
%

\institute{Lund Observatory, Division of Astrophysics, Department of Physics, Lund University, SE-22100 Lund, Sweden\\
              \email{dainis@astro.lu.se}
\and
     Zentrum f\"{u}r Astronomie der Universit\"{a}t Heidelberg, Landessternwarte, K\"{o}nigstuhl 12, DE--69117 Heidelberg, Germany\\
              \email{hludwig@lsw.uni-heidelberg.de}
\and
	Leibniz-Institut für Astrophysik Potsdam, An der Sternwarte 16, DE-14482 Potsdam, Germany\\
       	\email{msteffen@aip.de}
 \and
       Instituto de Astrofísica de Canarias, C/ Vía Láctea s/n, E-38205 La Laguna, Tenerife, Spain 	
\and
	Texas Advanced Computing Center, The University of Texas at Austin, Austin, TX 78758, USA\\}

\date{Received February 19, 2026; accepted April 7, 2026}

\abstract
  % context heading (optional)
   { Finding low-mass planets around solar-type stars requires to understand the physical variability of the host star, which greatly exceeds the planet-induced radial-velocity modulation.  Different solar photospheric absorption lines have slightly disparate responses to stellar activity, which should permit to disentangle wavelength shifts induced by exoplanets from those originating in stellar atmospheres. }
  % aims heading (mandatory)
   {Changing area coverage of magnetic active-region granulation (faculae and plage) causes radial-velocity fluctuations of the disk-integrated solar spectrum, whose precise modeling requires active-region spectral line profiles.  Hydrodynamic 3D modeling of granulation in magnetic fields extends previous non-magnetic studies, revealing different line profiles and altered convective velocity shifts. }
  % methods heading (mandatory)
   {Different types of lines in the visual and near infrared are examined in synthetic hyper-high resolution spectra ($\lambda$/$\Delta\lambda$ $\sim$900,000), comparing non-magnetic areas with those with strongly magnetic (240 mT = 2400~G) granulation.}
  % results heading (mandatory)
   {Magnetic fields inhibit convective motions, decrease the energy flow, produce more symmetric lines, and remove the common blueshift with its familiar `C'-shape bisectors.  Unexpectedly, magnetic granulation displays convective redshifts.  Their origin is traced to contributions from small areas, where hot and bright down-moving elements are created through shocks and adiabatic compression when rising gas is forced over into magnetically channeled downflows. }
  % conclusions heading (optional)
   {Understanding line formation in also stellar active regions is needed to simulate full-disk spectra toward exoEarth detections.  Detailed shapes of spectral lines carry significant information, suggesting that hyper-high spectral resolution may ultimately be required. }


\titlerunning{Solar photospheric spectrum microvariability. III}
\authorrunning{D. Dravins et al.}
\maketitle 

\nolinenumbers

\begin{figure*}

\centering
 \includegraphics[width=18cm]{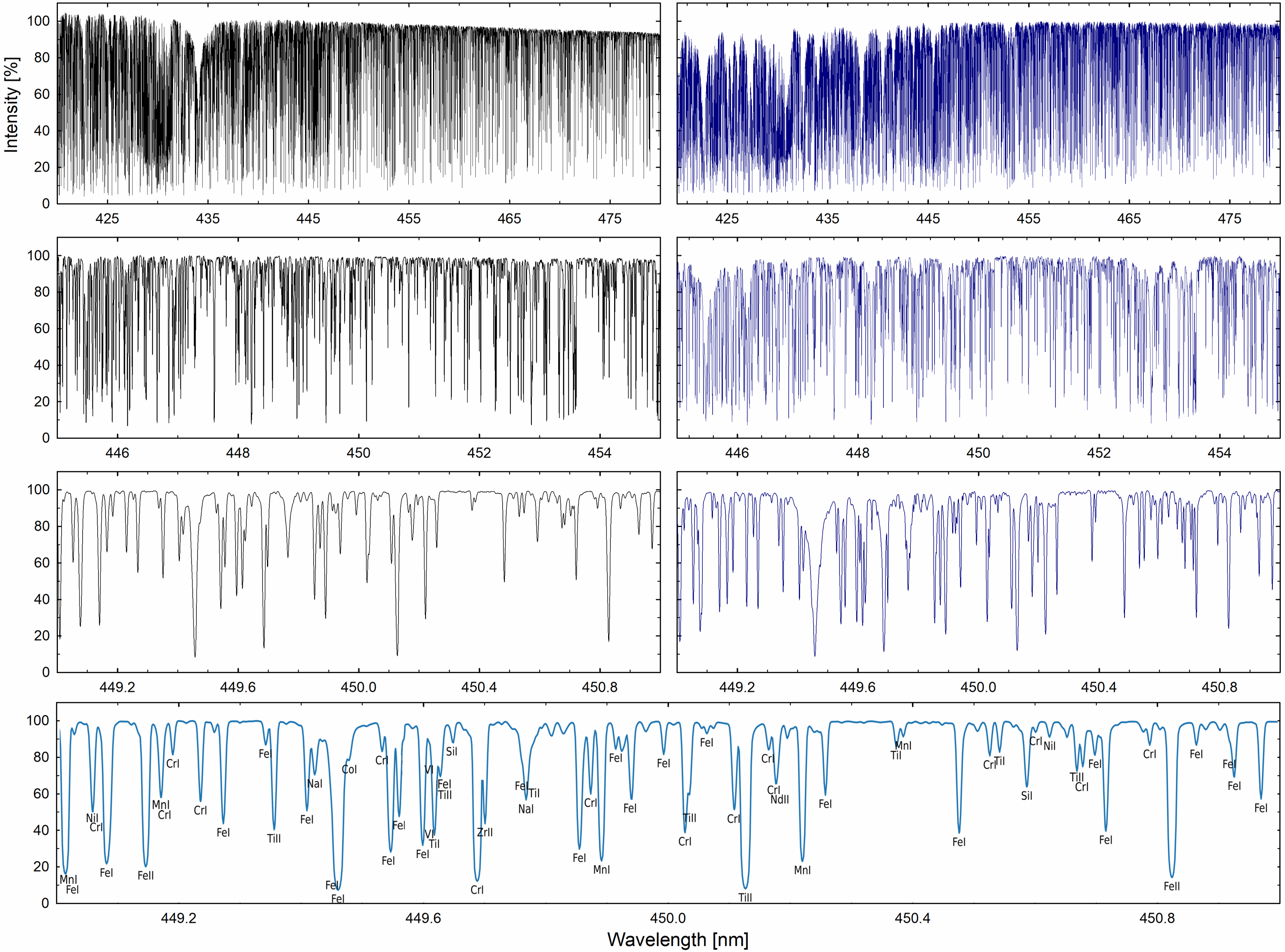}
     \caption{Example of synthetic spectra at the solar disk center, ${\mu}$=1, for the non-magnetic (left) and the magnetic 240 mT (2400~G) simulation, expanded in wavelength from top down. Each section is normalized locally while the lower frame shows line identifications.  From that region, the lines of \ion{Fe}{ii} 449.1405 and \ion{Fe}{ii} 450.8288 nm were selected for detailed study (Fig.\ \ref{fig:feiilines} below).  The wavelength scale denotes values in air.  }  
\label{fig:spectrumsample}
\end{figure*}

\section{Introduction}

Efforts are underway toward enabling the detection of `exoEarths', i.e., planets of about one Earth mass, in about one-year orbits around solar-type stars.  The most promising method appears to be the radial-velocity one, although the signal is minuscule: for an Earth-mass planet orbiting a solar-mass star in a one-year orbit, at most only 10~cm\,s$^{-1}$ \citep[e.g.,][]{halletal18}.  Although hardware and software developments now approach such precisions  \citep[e.g.,][]{artigauetal22, blackmanetal20, crassetal21, cretignieretal21, fischeretal16, fordetal24, guptabedell24, osullivanaigrain24, rackhametal23, salzeretal25, wilkenetal12, zhaoetal26}, such an exoplanet signal is overshadowed by much greater intrinsic stellar variability.  To reach adequate sensitivity thus requires to understand the complexities of stellar atmospheric dynamics and spectral line modulation.  A step toward exoEarth detection can be to identify dissimilar spectral lines (e.g., strong or weak, neutral or ionized, high or low excitation, atomic or molecular, short or long wavelength, magnetically sensitive or not), with disparate responses to stellar activity, to disentangle wavelength shifts induced by exoplanets from those originating in solar-type atmospheres.  

On the theoretical side, simulations with time-dependent 3D magnetohydrodynamics now provide quite realistic descriptions of solar photospheric spectral-line forming regions.  Using the output from such simulations as arrays of time-dependent 3D atmospheres, complete stellar spectra can be computed, incorporating full transition databases such as VALD\footnote{The Vienna Atomic Line Database of atomic and molecular transition parameters of astronomical interest.} \citep{heiteretal15, ryabchikovaetal15}, while sampling wavelengths with a hyper-high spectral resolution, $\lambda$/$\Delta\lambda$\,$\sim$1,000,000\footnote{The term `hyper-high' is used to denote spectral resolutions $\lambda$/$\Delta\lambda\gtrsim$\,10$^6$ since `ultra-high' is already in common use to describe spectrometers with the much lower resolutions of ´only' $\sim$200,000.} \citep{chiavassaetal18, dravinsetal21a, dravinsetal21b, dravinsludwig23}. 

Non-magnetic granulation covers most of the solar surface and provides most of the solar irradiance.  In Paper~I \citep{dravinsludwig23}, sequences of synthetic spectra computed from corresponding 3D simulations were examined to identify patterns of short-term variability in the quiet Sun.  The second largest contribution to solar irradiance comes from plage and faculae areas of magnetic granulation.  Magnetic fields disturb and dampen the convective velocity patterns, cause different asymmetries of the line profiles and modify wavelength shifts.  The varying area coverage of magnetic granulation during the solar cycle appears to be the main driver for longer-term changes in solar apparent radial velocity \citep{lakelandetal24, meunieretal10a, meunieretal10b, meunieretal24}.  While contributions from sunspots also affect spectral line shapes \citep{komorietal25}, their fraction of the solar photospheric flux is modest.

\section{Appearance of magnetic granulation on the Sun}

Much of the small-scale magnetic fields across granulation outside sunspots is outlined by the bright network.  This magnetic network appears bright in various spectral regions, in particular in molecular and other temperature-sensitive lines.  Among these, observations and modeling in the {\it{G}}-band are particularly extensive \citep[e.g.,][]{bergeretal04, kuridzeetal25, rouppevandervoortetal05}.  Similarly to other stronger lines, the network is also visible in the wings of \ion{Na}{i} D$_1$ \citep{jessetal10, keysetal13, keysetal26}. At high spatial resolution, it resolves into solar filigree (sometimes called `bright points', although not really point-like), occupying spaces between granules. The corresponding appearance in the chromosphere is more smeared out, apparently reflecting the expansion of magnetic flux into higher layers. 

A review of magnetic fields in the quiet Sun is given by \citet{bellotrubioorozcosuarez19}.  Different spectral lines have varying sensitivities to the magnetic field strength \citep{sinjanetal24, quinteronodaetal21}, while near-infrared lines might be especially valuable for mapping fields in the deeper photospheric layers \citep{hahlinetal23, laggetal16}.  Values on the order of 160 mT (1600~G) seem characteristic for actual intergranular field strengths \citep{bellotrubioorozcosuarez19, dewijnetal09, salhabetal18, stein12}, while fields in sunspots may extend to 400 mT and beyond.

Convective blueshifts are suppressed in magnetically disturbed granulation and line profile asymmetries and wavelength shifts are observed to gradually change when approaching and entering plage regions of progressively higher magnetic activity, as delineated by increasingly bright flux in the \ion{Ca}{ii}~K line or by greater filling factors for the magnetic flux (e.g., \citealt{brandtsolanki90, cavallinietal85, cavallinietal86, cavallinietal88, cavallinietal89}). For strong lines formed high in the atmosphere, even inverted line asymmetries are observed, although their origins probably lie in more complex chromospheric dynamics (e.g., \citealt{uitenbroek06}).

\begin{figure}
\centering
 \includegraphics[width=\hsize]{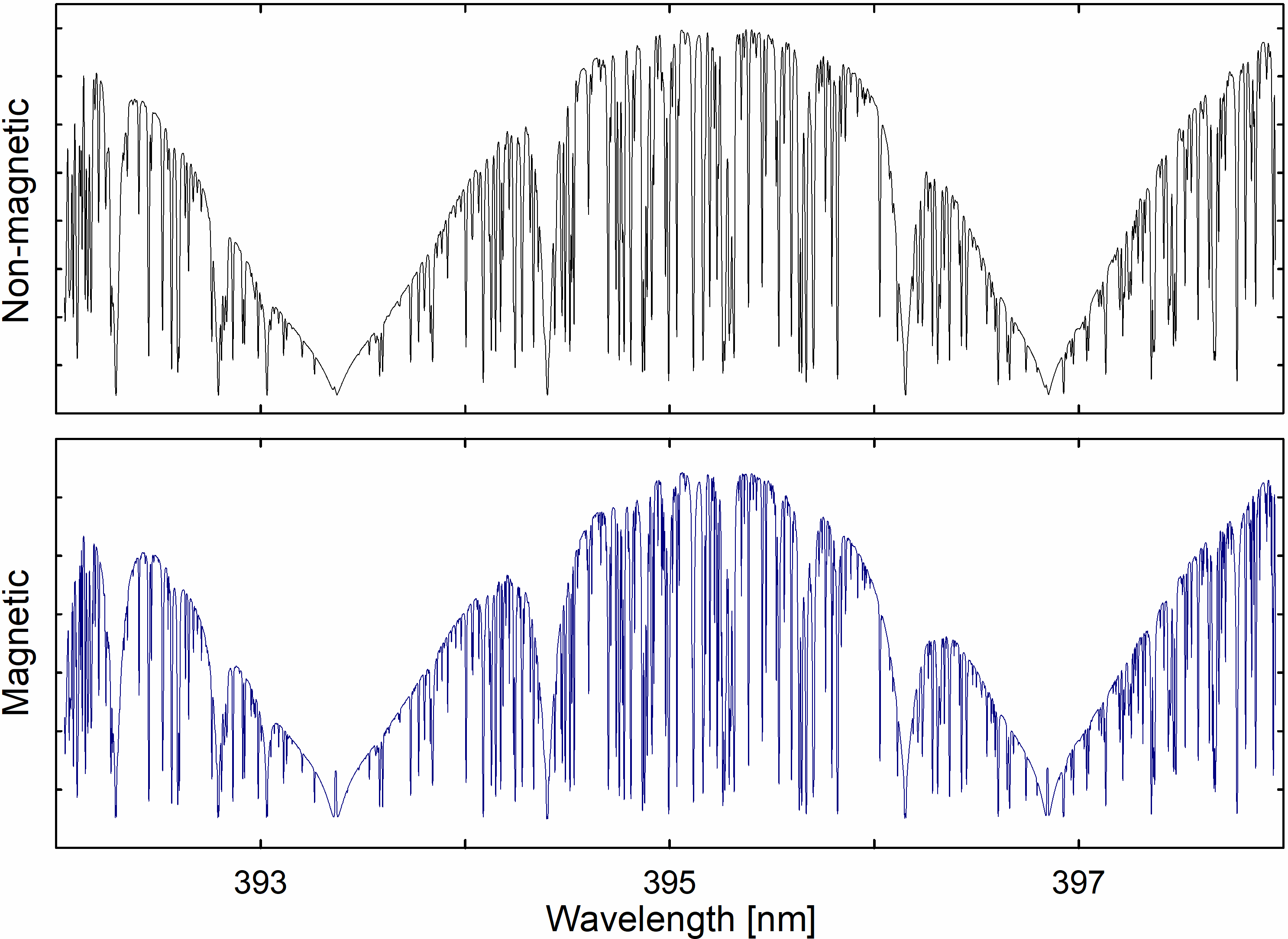}
     \caption{Synthetic spectra in the \ion{Ca}{ii}\,H \& K line region: non-magnetic (top) and 240 mT (2400~G) magnetic models.  (However, the bottoms of these very strong lines are not precisely reproduced by present modeling.)  From this particular region, \ion{Fe}{i} 393.2627 and \ion{Fe}{i} 396.7421 nm lines, superposed onto the extended and sloping absorption wings of the K and H lines, were selected for examination (Fig.\ \ref{fig:feicaii} below).  These spectra are for the integrated solar disk. }  
\label{fig:hklines}
\end{figure}

\section{CO\,$^5$BOLD models of the solar photosphere}

In Paper~I \citep{dravinsludwig23} and earlier, spectra from 3D simulations of the quiet non-magnetic solar granulation were examined.  We now turn to magnetic variants.  A sequence of magnetohydrodynamic CO\,$^5$BOLD solar models were computed with the same resolution as in the non-magnetic case by modeling a Cartesian `box-in-a-star' of 140\,$\times$\,140\,$\times$\,150 grid cells, with horizontal stepsizes $\Delta$x = $\Delta$y = 40 km and vertical $\Delta$z = 15 km.    For details about such computations, see \citet{freytagetal12} and \citet{tremblayetal13}.

These magnetic simulations represent a range of average flux densities $\langle$B$_z$$\rangle$ from zero to 320 mT = 3200~G \citep{ludwigetal23}.  The total net magnetic flux was set by initial conditions and preserved during the modeling sequence, with boundary conditions forcing fieldlines to pass orthogonally through the top and bottom boundaries of the simulation volumes.  Thus, the horizontally averaged magnetic flux density remains constant during the evolution of the magnetized flow.  To generate the initial model, a unipolar and homogeneous vertical magnetic field was added to a non-magnetic temporal snapshot.  After injecting such a magnetic flux in 3D atmospheric structures, its development can be followed with magnetic flux concentrating around intergranular lanes, the magnetic forces being balanced by gas pressure and dynamic effects, resulting in a somewhat disturbed and disrupted granulation structure.  Radiative transfer is treated in 12 opacity bins and the simulations are advanced by an HLL (Harten-Lax-van Leer) solver, stabilized with additional turbulent viscosity and entropy diffusion.  Irrespective of the magnetic flux, the heat content of the inflowing material was kept constant.  By limiting the maximum Alfvén speed, all simulations could be run with a uniform time step of $\Delta$t = 0.2 s, even for the highest magnetic field strengths.  Each simulation covers 12.5 hours of solar time, including initial relaxation.  This time is substantially longer than for the non-magnetic models, permitting to follow the slower evolution of magnetic features. The vertical extent of the simulation box is 2250 km, and its top layer corresponds to Rosseland optical depth log\,{$\tau$}= $-$6.5 in the non-magnetic case and log\,{$\tau$}= $-$6.6 for the 240 mT magnetic model.  For details, including synthetic surface images for different magnetic flux levels and at different limb angles, see \citet{ludwigetal23}.  

Somewhat similar surface magnetoconvection has been modeled by \citet{beecketal15a}, \citet{bhatiaetal22, bhatiaetal26}, \citet{khomenkoetal18}, \citet{norrisetal23}, \citet{salhabetal18}, \citet{shuklaetal26}, \citet{vogleretal05}, with spectral features computed by \citet{beecketal15b, smithaetal21}, and others.  The radiative properties of these MHD simulations appear to be qualitatively consistent and indicate that, as long as the magnetic field is sufficiently weak, the radiative flux is enhanced by the presence of magnetic flux concentrations but reverses somewhere around $\langle$B$_z$$\rangle$ = 50 mT.  Gradually, darker features become increasingly prominent in the simulations with larger magnetic fluxes.  For our strongest fields, $\langle$B$_z$$\rangle$ = 320 mT, the structures start to resemble sunspot umbrae; the magnetic flux density can locally reach very high values, with convection now restricted to narrow, almost field-free plumes of high upward velocities.  In the synthetic continuum intensity images, these plumes resemble bright and moving `umbral dots’, often with the visual appearance of `coffee beans': \citet{ludwigetal23}, and Fig.\ \ref{fig:3dmodel} below.  

A significant reduction of the bolometric radiative flux follows in such stronger fields and with even stronger flux concentrations, surface convection is inhibited and spots start to develop.  With appropriate boundary conditions, models of umbrae and penumbrae structures can then be constructed \citep{panjaetal20, schusslervogler06, smithaetal25}.   The more moderate field strengths treated here reduce the overall heat flux, lower the average surface temperature, and cause many spectral lines to strengthen, akin to stars somewhat cooler than the Sun (Fig.\ \ref{fig:spectrumsample}), even if they may weaken inside specific magnetic features.  However, the present simulations are not to model specific magnetic elements, but rather to estimate how radial velocities and line profiles are affected when an extended area of granulation is modified by the presence of magnetic flux.   

\subsection{Limitations of magnetic models}

Non-magnetic granulation can be modeled from `basic principles', using the fundamental parameters of stellar temperature, surface gravity and chemical composition.  Different models may apply different physical, mathematical or numerical approximations, and the ensuing spectral synthesis may be carried on different levels of sophistication, but there are no further freely adjustable physical parameters.  However, magnetic granulation introduces additional degrees of freedom, especially if modeling an entire star.  Not only can the area coverage and latitudinal and longitudinal surface distributions be different, but also the distribution of strengths within the magnetic flux concentrations.  For the spectral synthesis, an additional complication (or opportunity) comes with the possible treatment of magnetically sensitive lines.  Thus, results from magnetic granulation cannot be expected to be singularly unique but may have to be evaluated for a series of different magnetic field strengths and geometries, or one can look for characteristic line-profile signatures that differ between magnetic and non-magnetic regions.  Nonmagnetic simulations have in the past been verified against solar surface images, movies, and spectra, enabling successive fine-tuning of such models, but corresponding tests for magnetically well-defined samples are scarce (also because their atmospheric structures may be at, or below observational resolution limits), and their models may still have room for improvement.

\begin{figure*}
\sidecaption
 \includegraphics[width=12.9cm]{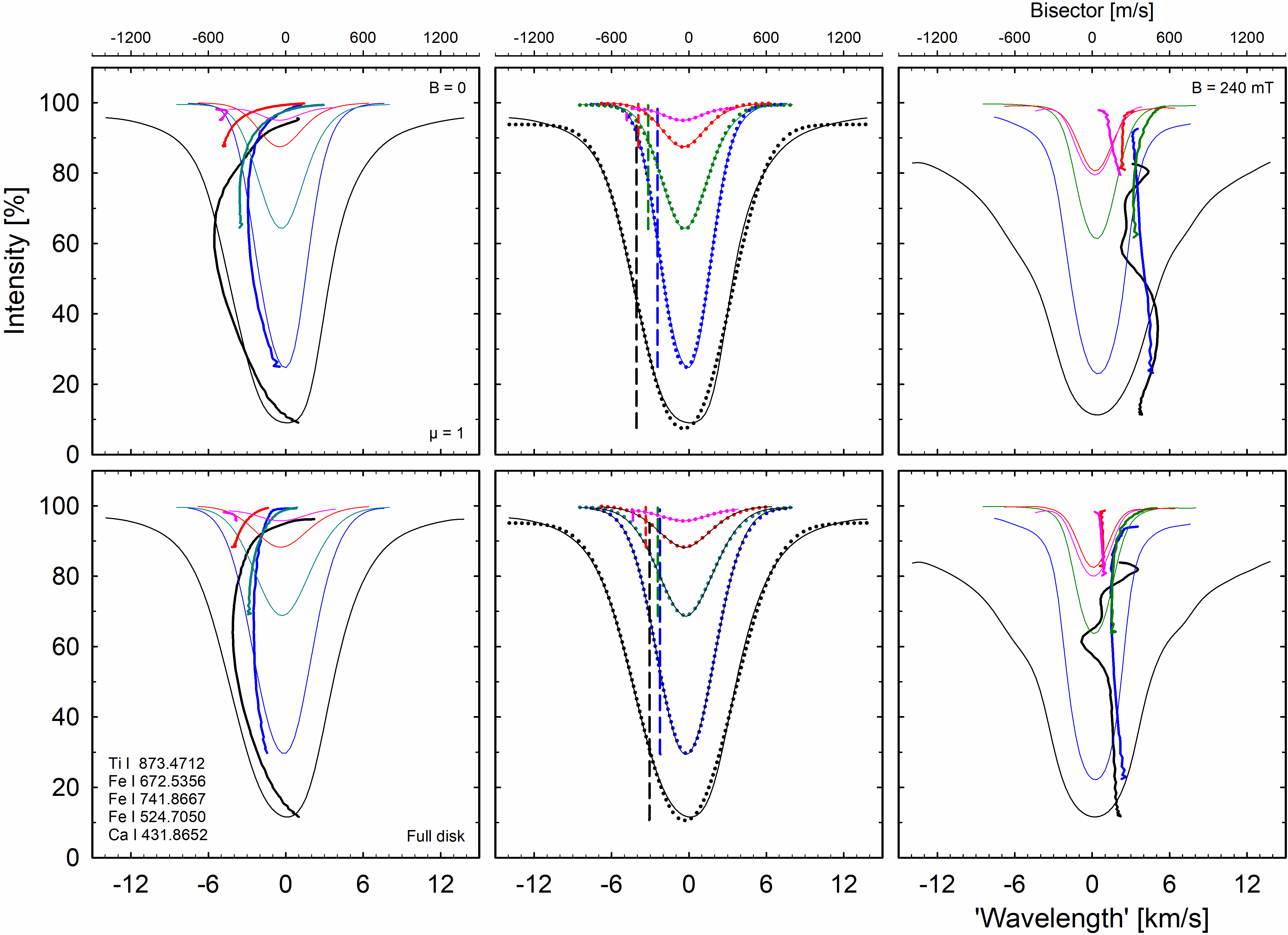}
     \caption{Modeled and fitted line profiles.  Left: Profiles and bisectors for differently deep absorption lines, as obtained from time-averaged 3D simulations of non-magnetic granulation.  Wavelengths are in velocity units relative to each line's laboratory wavelength.  At top, the bisector scale is expanded tenfold. Central column shows the same profiles from these simulations (dotted), now together with their fitted profiles (solid) using a five-parameter Gaussian-type function.  Since the fitted profiles are symmetric, their dashed bisectors are vertical limes, defining each line's average radial velocity.  The right-hand column shows the same lines in the 240 mT (2400 G) magnetic simulation. The upper row gives data for solar disk center, ${\mu}$ = 1, the lower one for integrated sunlight.  The lines are listed at bottom left, in order of increasing line-depth.  Bisector wiggles for the strongest magnetic line are caused by blends in its broad wings.  Effects of solar rotation and gravitational redshift are not included. } 
\label{fig:linedepths}
\end{figure*}

\begin{figure*}
\sidecaption
 \includegraphics[width=12.9cm]{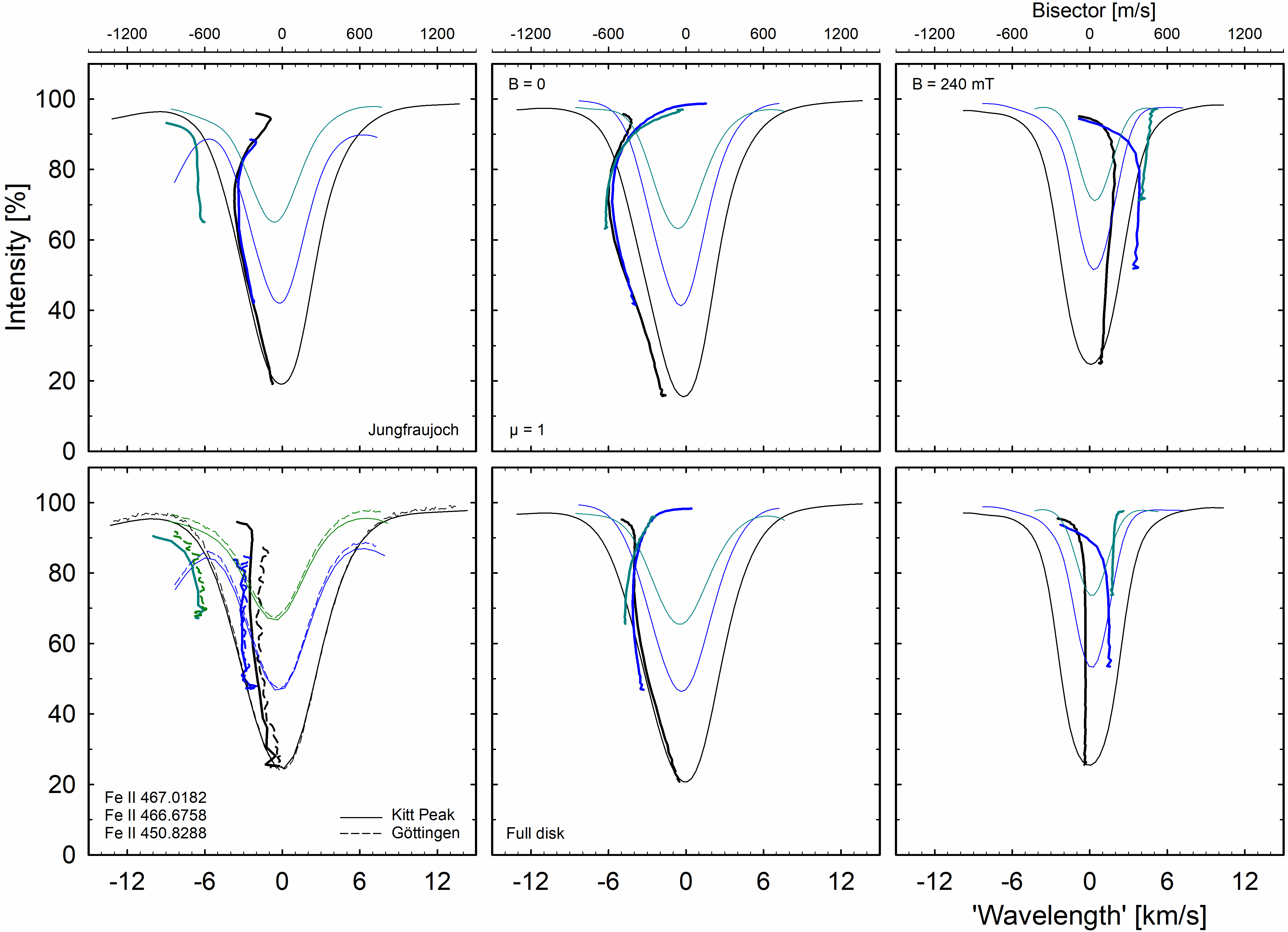}
     \caption{Line profiles and bisectors for differently strong \ion{Fe}{ii} lines.  The leftmost column shows observed data from the Jungfraujoch solar disk-center atlas (top) and the integrated solar flux atlases of Kitt Peak and Göttingen (bottom).  Profiles from the non-magnetic (center) and magnetic 240 mT = 2400~G simulations (right) are shown for disk center ${\mu}$ = 1 (upper row) and for integrated sunlight of a non-rotating Sun (lower).  The wavelength scales are relative to the laboratory values used as input to the simulations while the observational atlases are affected by solar gravitational redshift of 635 m\,s$^{-1}$ and likely errors in laboratory wavelengths.  }  
\label{fig:feiilines}
\end{figure*}

\subsection{Synthetic spectra from MHD simulations}

For these initial examinations of synthetic line profiles, a strong-field magnetohydrodynamic model with 240 mT = 2400~G was selected (referred to below as `magnetic granulation').  Such a field strength can be seen to represent the strongest-field granulation in active plage or facular areas or perhaps umbral dots, with a flux density in-between the field-free situation, and a fully developed sunspot umbra. 

From the CO\,$^5$BOLD simulation volumes, spatially and temporally averaged line profiles were computed in LTE (local thermodynamic equilibrium) using the ASS$\epsilon$ET code \citep{koesterkeetal08}.  Calculations were made for the full spectral range of 120--3000 nm, and for 20 largely uncorrelated snapshots in time during the simulation. The spectra incorporate practically all relevant atomic and molecular transitions with a wavelength sampling corresponding to the hyper-high resolving power of $\lambda$/$\Delta\lambda$ $\sim$900,000.  For different elevation and azimuth angles, the spectra were computed on a grid of 47\,$\times$\,47 points (each of horizontal extent 120\,$\times$\,120\,km$^{2}$, subsampled from 3\,$\times$\,3 pixels in the hydrodynamic simulation).  Integrated sunlight is computed by appropriately interpolating and summing spectra from different $\mu$-values, somewhat analogous to how full-disk signatures were synthesized by \citet{ceglaetal19}, \citet{frameetal25} or \citet{palumboetal22}.  The magnetic structures alter the shapes of spectral lines \citep{holzreutersolanki12, holzreutersolanki15, holzreuteretal25} while pairs of lines with disparate Land{\'e} $\varg$$_{\textrm{eff}}$-factors, and thus different magnetic sensitivities, should be able to more directly diagnose field properties \citep{solanki93, smithasolanki17}.  However, in the current calculations, Zeeman broadening or splitting is not included, and neither are solar rotational broadening or gravitational redshift (Figs.\ \ref{fig:spectrumsample} and \ref{fig:hklines}).  For later studies, especially concerning specific magnetic structures, this should permit to separate the indirect magnetic effect on the spectral lines as studied here (via the altered flow field) from the direct magnetic effect (via Zeeman broadening).  

Precision radial velocities are measured as some statistical average over many thousands of photospheric lines, where the different behavior between individual single lines normally is not measurable.  Larger line-groups may realistically be segregated against ‘simple’ parameters such as line-depth or wavelength region and lines representing such subgroups were selected for the present study.  However, given that the many lines entering radial-velocity measurements have varied dependences on Zeeman signatures, non-LTE sensitivities, etc., the selected ones should be seen as generic representatives for their particular groups of line-strength and wavelength region.  In particular, by being largely unblended, they should more clearly show how line shapes are molded and shaped by the altered gas flow patterns and temperature structures in magnetically modified granulation. 

The accuracy of synthetic spectra is ultimately limited by the quality of atomic and molecular data.  Current spectra include data from more than half a million atomic lines, compiled from a multitude of sources with sometimes variable accuracy.  The input wavelengths for the spectral synthesis are specified in nanometers to four decimal places.  At $\lambda$\,500 nm, the final decimal corresponds to a velocity of 60~m\,s$^{-1}$, which is one limiting parameter in comparing modeled line shifts to observed ones.  Even if the laboratory values were perfectly known, they are not specified to closer than these four decimal places and, in any case, they typically have errors of comparable magnitude, contributing further noise between synthetic and observed velocity shifts.  (However, differential shifts within synthetic spectra, are not affected by such errors, except for higher-order effects due to possibly imprecise wavelength superpositions of blending lines).  The absolute wavelength shifts discussed below, refer to the nominal laboratory values. 

Line-shape and wavelength variations between differently magnetic regions have been studied in the past.  However, such observations have been of relative wavelength shifts or between different positions in the line bisectors. Even with a very accurate spectrometer calibration, any deduction of absolute wavelength shifts depends on the accuracy of the laboratory wavelengths.  One main difference to the present work is that, by using numerically exact values for the transition wavelengths in the synthetic spectra, absolute lineshifts can be deduced relative to those values, even if they would not precisely agree with the true (but not perfectly known) laboratory values. 

A general limitation of photospheric modeling is that its vertical extent does not fully embrace the formation heights of stronger lines.  Also, boundary conditions might cause spurious temperature fluctuations near the upper boundary, contributing to emission artifacts in the cores of the strongest lines.  Thus, lines formed in the uppermost photosphere or lower chromosphere cannot be expected to be precisely reproduced.  Nevertheless, we will still examine the behavior of some strong lines such as the \ion{Mg}{i} triplet, the \ion{Na}{i} D lines, and the infrared \ion{Ca}{ii} triplet, because -- even if their cores cannot be faithfully modeled -- their largely photospheric flanks and wings should still be approximately reproduced and might confirm trends seen in other lines when going toward higher atmospheric layers. 

\begin{figure*}
\sidecaption
 \includegraphics[width=12.5cm]{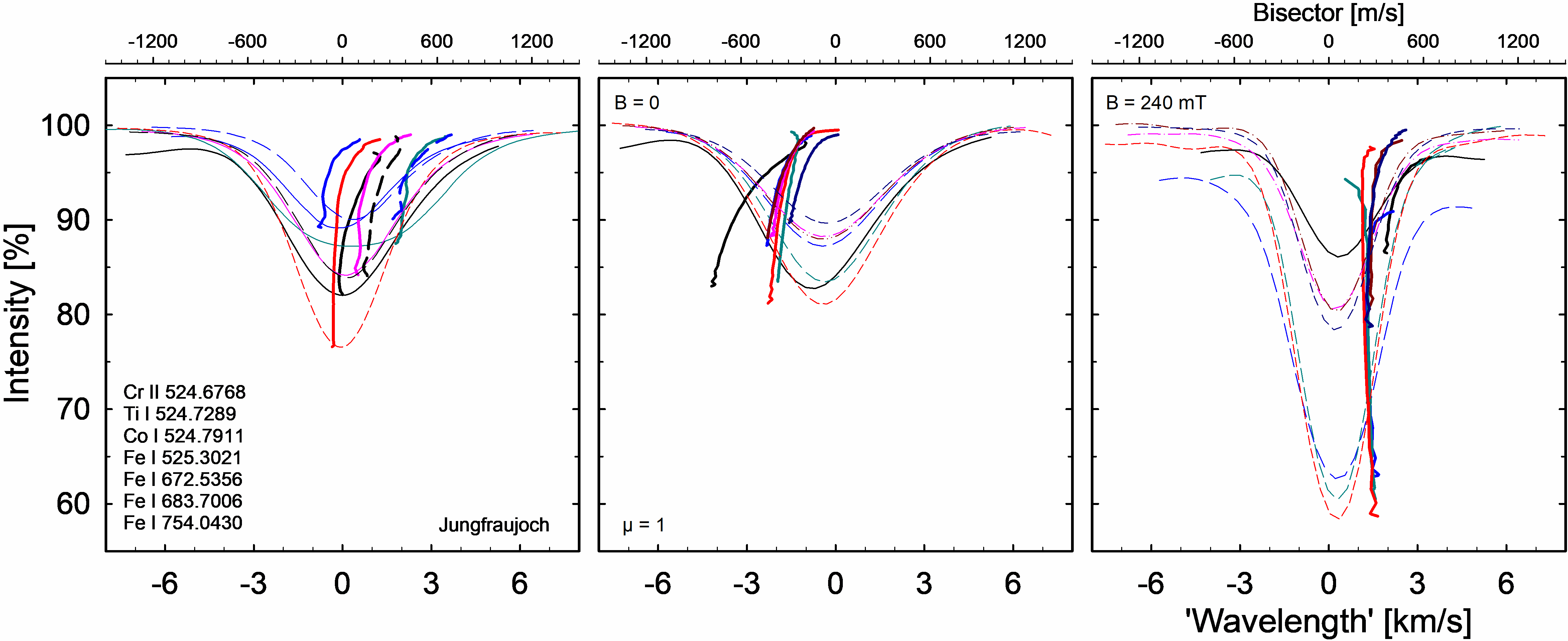}
     \caption{Weak lines at solar disk center, ${\mu}$ = 1. Left: line profiles and bisectors from the Jungfraujoch atlas.  Numerous lines of comparable strength show similar behavior in both the non-magnetic case (center) and in the 240 mT = 2400~G magnetic simulation (right).  Wavelengths in observational atlases are affected by the solar gravitational redshift of 635 m\,s$^{-1}$ and possible errors in laboratory wavelengths.}  
\label{fig:weaklines}
\end{figure*}

\begin{figure*}
\sidecaption
 \includegraphics[width=12.9cm]{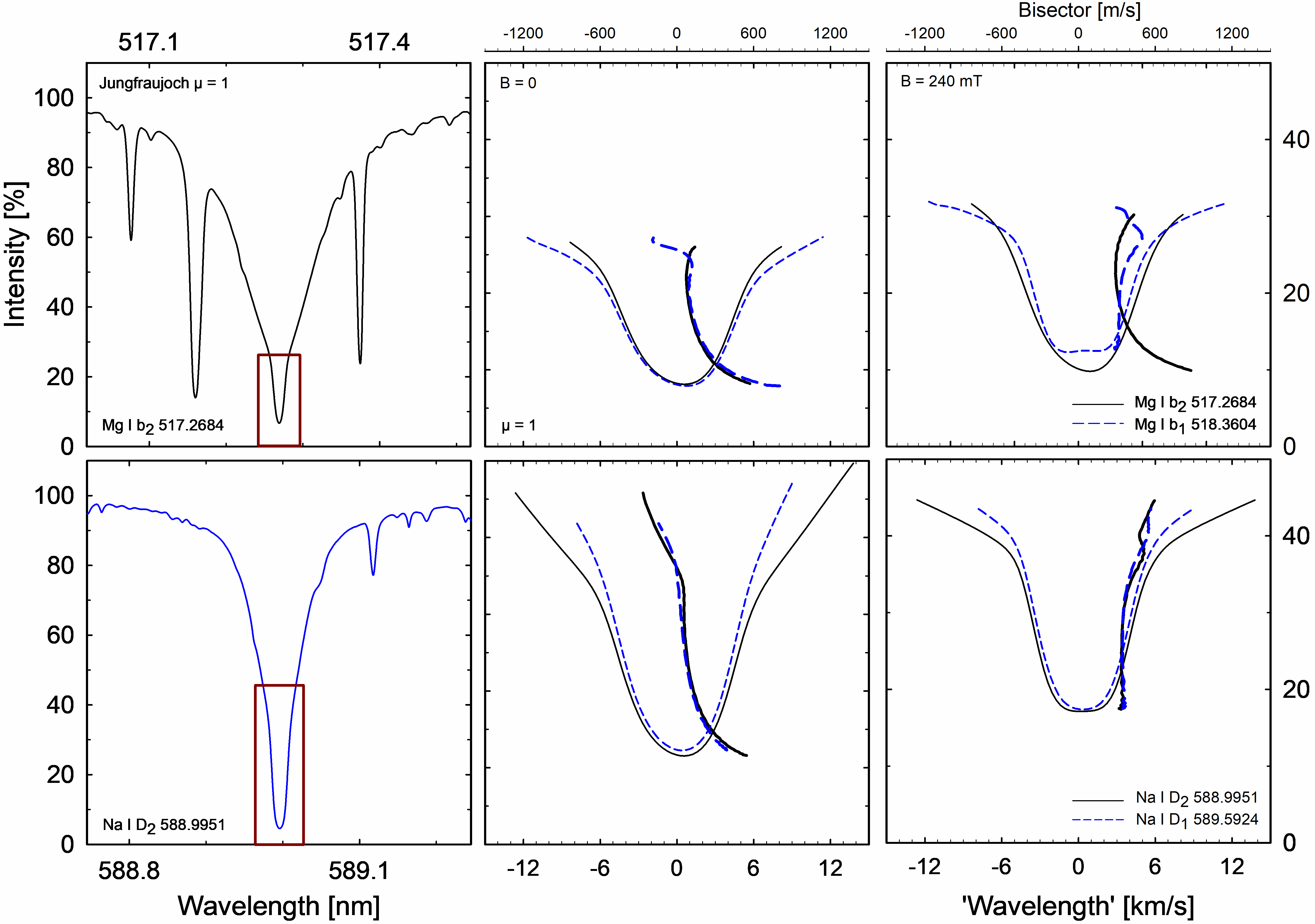}
     \caption{Very strong lines at solar disk center. Left: \ion{Mg}{i} b$_2$ and \ion{Na}{i} D$_2$ lines in the Jungfraujoch atlas.  Center: synthetic profiles and bisectors for their line cores (corresponding to the boxes at left), and also the related \ion{Mg}{i} b$_1$ and \ion{Na}{i} D$_1$ ones, for the non-magnetic simulation.  Right: the 240 mT = 2400~G magnetic case.  Effects of solar rotation and gravitational redshift are not included.}  
\label{fig:stronglines}
\end{figure*}

\begin{figure*}
\sidecaption
 \includegraphics[width=12.9cm]{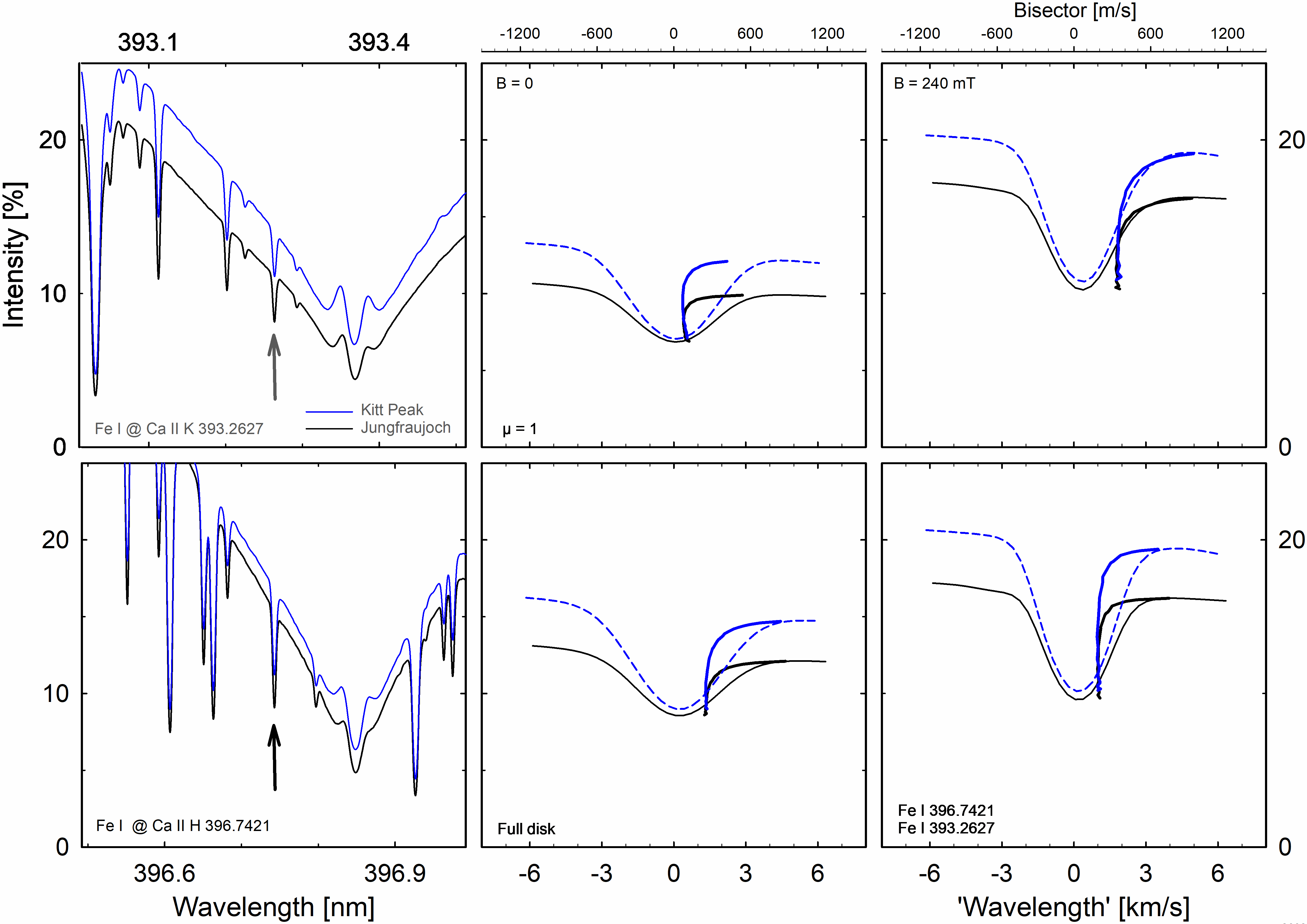}
     \caption{Lines in special positions display special behavior.  The weak \ion{Fe}{i} 393.2627 and \ion{Fe}{i} 396.7421 lines are superposed onto the extended absorption wings of the \ion{Ca}{ii} H\&K lines and get their formation heights lifted through combining their opacities with those of the H\&K line wings.  The leftmost column shows their appearance in solar spectrum atlases (Kitt Peak full-disk, Jungfraujoch disk center).  With vertical bisectors, the line cores are remarkably symmetric in both the non-magnetic (center) and magnetic 240 mT = 2400~G case (right). }  
\label{fig:feicaii}
\end{figure*}

\section{Selecting and fitting representative spectral lines} 

The spectral region selected was 350--1000 nm, spanning the observational regions for current visual and near-infrared radial-velocity instruments \citep[e.g.,][]{bouchyetal25,  doyonetal25, locurtoetal24, mayoretal03}.  About 50 lines of various strengths in different wavelength regions, that appeared largely unblended in both the nonmagnetic and magnetic spectra were selected for closer examination.  Given that Fe lines are ubiquitous throughout the spectrum and are commonly used for various atmospheric diagnostics and velocity shifts, most lines are \ion{Fe}{i} and \ion{Fe}{ii}, many the same as in Paper~I (Table \ref{table:linelist1}).  To include also a very broad photospheric line, a \ion{Ca}{i} was chosen (despite such a line no longer being fully unblended), and lines of varying strength from several other species are also included.  

All lines are asymmetric and variable during the simulation sequences. To practically examine dependences among various line parameters, the information in the line profiles needs to be parameterized.  Following various tests of line-fitting parameters, a suitable approximation was found (as in Paper~I) by fitting the synthetic line profiles with five-parameter Gaussian-like functions of the type $y_0 - a\cdot\exp[-0.5\cdot(|x-x_0|/W)^c]$, yielding $y_0$ for the continuum level, $a$ for the absorption depth, $W$ as a measure of the line width, $c$ for the degree of the line’s boxiness or pointedness, and $x_0$ for the `average' wavelength.  As illustrated in Fig.\ \ref{fig:linedepths}, this fitting provides one unique value for the radial velocity that refers to an average displacement of the line as a whole.  The exact value that a radial-velocity spectrometer will obtain, depends on the precise algorithm and type of weighting applied; however, the physical differences between different spectral lines are much greater than plausible differences in the definition of such a spectral-line `wavelength'.

Very strong lines from the \ion{Mg}{i} triplet, \ion{Na}{i} D lines, and the infrared \ion{Ca}{ii} triplet were treated somewhat differently with radial velocities computed separately for their line cores and flanks.  In the magnetic spectra, some of their cores start to exhibit irregular emission, which does not provide meaningful radial velocities, while the more photospheric line-flanks still might.

\section{Profiles and velocity shifts for different lines }

A general property of photospheric lines from quiet non-magnetic granulation is their asymmetry and convective blueshift, well studied both observationally and theoretically \citep[e.g.,][]{bergemannhoppe26, reinersetal16, sheminova22}.  A characteristic signature is the marching progression of ´C'-shaped bisectors for differently strong lines, with the weakest lines normally showing the largest blueshifts.  In Fig.\ \ref{fig:linedepths}, such a sequence of lines with gradually increasing strengths is assembled, showing this familiar bisector pattern, with amplitudes at disk center being somewhat greater than in integrated sunlight from the full disk.

\subsection{Lines of different strength}

The central frames of Fig.\ \ref{fig:linedepths} illustrate how this fitting procedure provides unique values for the average radial velocities of each line.  The synthetic profiles (dotted) are fitted with that five-parameter function, which is seen to very closely follow the synthetic ones, the only exception being slight differences in the core and wings of the very broad \ion{Ca}{i} line.  Since the fitted profiles are symmetric, their bisectors are vertical lines, whose positions now define each line's average radial velocity, as used in the later plots and tables.  All wavelength scales are relative to the database values for laboratory wavelengths used as input to the spectral synthesis and modeled lineshifts are thus not affected by possible errors in the laboratory values.  However, even if those lines are selected to be apparently unblended, some astrophysical noise is contributed from numerous weak blending lines: the input catalog has on order of 1,000 lines per each nm interval and every line will have at least slight contributions from also others. 

The right-hand frames of Fig.\ \ref{fig:linedepths} show the same spectral lines from the magnetic simulation.  In this case, their sequence of relative line-strengths does not change much but their bisectors do.  The lines largely lose their asymmetries with their bisectors tending to become straight vertical lines (disregarding the bisector wiggles for the strongest magnetic line, which are caused by blends appearing in its broad line wings).  Strikingly, the convective blueshift in non-magnetic granulation is now instead replaced by a redshift.  Analogous to the nonmagnetic case, this shift away from zero is more pronounced at disk center (top) than in integrated sunlight.

\subsection{Closely related lines}

Fig.\ \ref{fig:feiilines} attempts to examine these effects for three intermediate and strong lines selected to be very similar, except for their depths: \ion{Fe}{ii} $\lambda$\,450.8288 nm, $\chi$\,=\,2.86 eV; \ion{Fe}{ii} $\lambda$\,466.6758, $\chi$\,=\,2.83; \ion{Fe}{ii} $\lambda$\,467.0182 nm, $\chi$\,=\,2.58 eV.   Analogous to the non-magnetic case, the largest shifts for the magnetic profiles are seen in the weaker lines, while the bottoms of the strongest ones show only modest displacements.

Besides the synthetic data, observed spectra are shown (left) from the Jungfraujoch disk-center atlas  \citep[top;][]{delbouilleetal89} and from the integrated flux atlases of Kitt Peak \citep{kuruczetal84} and Göttingen \citep{reinersetal16}, illustrating both similarities and typical differences.  Wavelength shifts from the observational atlases refer to the same database values (to four decimal points in nm) as the synthetic lines, and are subject to round-off errors of the true wavelength to this fourth decimal place, imprecisions in laboratory wavelengths, possible inaccuracies in the atlas wavelength scales, a somewhat different pattern of weak blending lines (whose wavelengths or strengths may not perfectly coincide with those in synthetic spectra); affected by a slight line smearing by an extended spectrometer slit at solar disk center, full-disk line broadening due to solar rotation, and the solar gravitational redshift of 635 m\,s$^{-1}$.  The exact gravitational redshift value depends on the precise height of line formation \citep{ceglaetal12, lindegrendravins03}.\footnote{Sometimes discrepant numbers are quoted.  Gravitational redshift from the solar nominal radius to infinity equals 636.310 m\,s$^{-1}$; from the solar surface to a static location at 1 AU: 633.351 m\,s$^{-1}$; when measured on the Earth moving with its average orbital speed, a transverse Doppler redshift of 1.481 m\,s$^{-1}$ enters, while the Earth’s own gravitational potential adds a blueshift of $\sim$0.209 m\,s$^{-1}$, for a total of $\sim$634.623 m\,s$^{-1}$. } 

\subsection{Very weak lines}

Fig.\ \ref{fig:weaklines} with seven lines from the disk center, illustrate the degree of similarities between different weak lines.  At left are observed profiles from the Jungfraujoch atlas, and then the profiles and bisectors for the nonmagnetic and magnetic models.  Lines of different species, and in different regions of the spectrum, display quite similar behavior.  The observed \ion{Co}{i} line is broader than the others, most probably due to its hyperfine structure, an effect not treated in the synthetic spectra.  Data for the full disk (not shown) are qualitatively similar.

\subsection{Very strong lines}

Fig,\ \ref{fig:stronglines} shows samples of the strongest lines at solar disk center (again, full-disk data are not much different).  Left column displays \ion{Mg}{i} b$_2$ and \ion{Na}{i} D$_2$ profiles in the Jungfraujoch atlas, center and right their modeled nonmagnetic and magnetic line-core profiles, also the very similar \ion{Mg}{i} b$_1$ and \ion{Na}{i} D$_1$ lines are plotted.  Due to the presence of blends across these broad lines, bisectors are fitted only to their deep cores (note change of intensity scales).

The limited applicability of photospheric granulation models to the detailed formation of very strong lines in the upper photosphere (with contributions from also the low chromosphere) was noted above.  For these types of lines, additional issues arise because of likely non-LTE effects.  The green magnesium triplet of \ion{Mg}{i} b{$_1$} $\lambda$\,518.3, b{$_2$} $\lambda$\,517.2, and b{$_3$} $\lambda$\,516.7 nm has coupled atomic energy levels such that these are transitions between one shared upper level and three different lower levels, contributing to the complexity of their formation.  For a discussion of their formation and references, see  Paper~II \citep{dravinsludwig24}.   

Other strong lines behave rather similarly to the green Mg triplet.  The two strong \ion{Na}{i} D$_1$ and D$_2$ lines are used as diagnostics for the upper photosphere and the lower chromosphere (usually, the weaker and less blended \ion{Na}{i} D$_1$).  The infrared \ion{Ca}{ii} triplet lines at $\lambda$\,849.8, $\lambda$\,854.2, and $\lambda$\,866.2 nm reflect activity in stars and their behavior in the current modeling largely follows \ion{Mg}{i} b and \ion{Na}{i} D.  However, in the magnetic model, spurious emission develops in the cores of these lines, limiting sensible radial-velocity determinations to the line flanks only (Table \ref{table:linelist1}).  Further details for these stronger lines were discussed in Paper~II, in connection with observations of their longer-term variations in the spectrum of the Sun seen-as-a-star, comparing to the classic chromospheric indicators of \ion{Ca}{ii}~H\,\&\,K.

\begin{figure}
\centering
 \includegraphics[width=7cm]{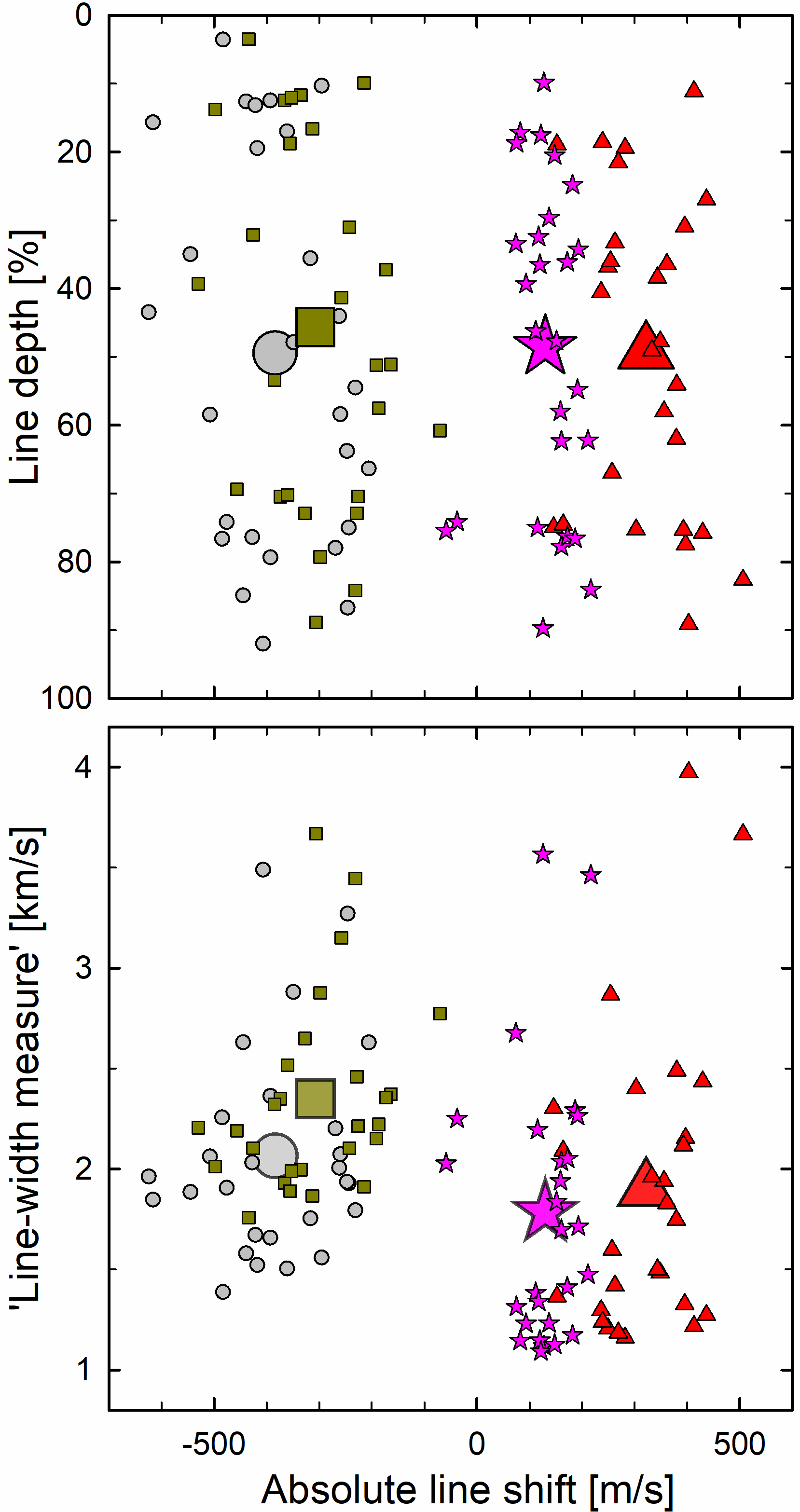}
     \caption{Radial-velocity shifts for ordinary lines (Table \ref{table:1}).  Gray circles are lines from the non-magnetic simulation at solar disk center; dark squares for the full solar disk.  Red triangles are from the magnetic model at disk center; purple stars are magnetic full disk.  Small symbols denote specific spectral lines; large symbols their total average.  Line depth denotes absorption in units of the local continuum; weak lines thus carry small numbers.  Each line's average wavelength and line-width measure is obtained from fitting the synthetic profiles to a symmetric five-parameter Gaussian-type function.  The modeled lines are not affected by either solar rotation or gravitational redshift. }  
\label{fig:lineshiftsall}
\end{figure}

\begin{figure*}
\centering
 \includegraphics[width=18cm]{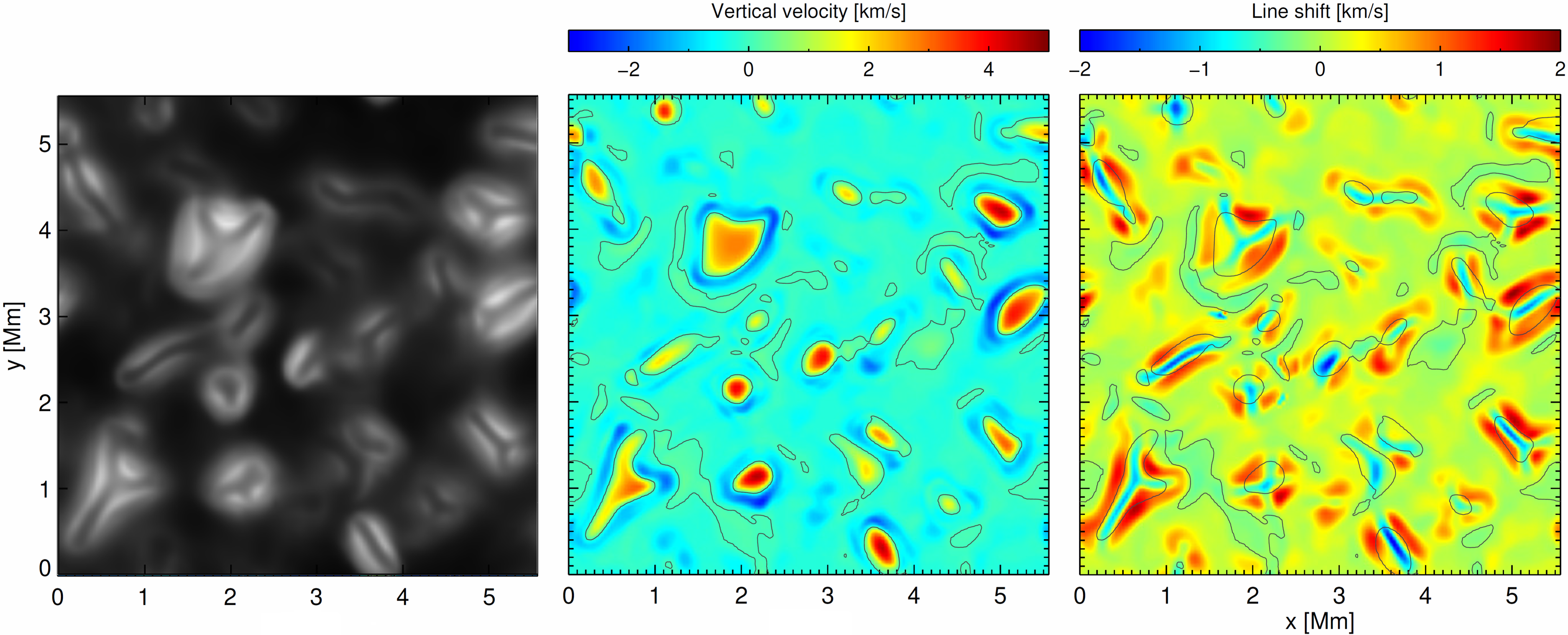}
     \caption{Brightness, gas velocities, and radial velocities in one temporal snapshot of the magnetic 3D simulation for 240 mT = 2400~G.  Left: Continuum intensity at solar disk center (${\mu}$ = 1) near \ion{Fe}{i} 683.7 nm.  Center: Vertical gas velocity (measured positive rising upward in the atmosphere) over a horizontal plane near Rosseland optical depth unity. Thin black lines depict the {\it{v}}$_z$ = 0 contour separating up- and downflows.  Right: Spectroscopic line shifts for \ion{Fe}{i} 683.7 nm (measured positive redshifted if receding downward into the atmosphere, thus inverting the color scale). }  
\label{fig:3dmodel}
\end{figure*}
 
\section{Lines in particular positions}

Some spectral lines combine with others, such as weak photospheric lines superposed onto the extended flanks of the very broad \ion{Ca}{ii} H\&K lines.  Although such lines would in isolation be ordinary ones, their formation now is affected by line-wing opacities or radiative transitions of the host line, leading them to sample different atmospheric conditions and -- under the combined opacities with its host line -- lifting its formation level into higher atmospheric layers.  Already long ago, such a behavior was discovered for the hydrogen Balmer H${\varepsilon}$ line in the longward wing of \ion{Ca}{ii}~H \citep{wilson38, hinkleetal00}.  Balmer lines normally are absorption features but H${\varepsilon}$ appeared as a chromospheric-type emission in the spectrum of the cool giant star Arcturus, since seen in numerous other stars.  On the Sun, high-resolution H${\varepsilon}$ spectroheliograms reveal a reversed granulation pattern, indicating enhanced formation heights \citep{krikovaetal23}.  Related solar phenomena include emission lines from rare earth elements and from \ion{Fe}{ii}, appearing in the \ion{Ca}{ii} H\&K line wings, pointing to non-LTE effects. 

The solar H${\varepsilon}$ is a broad and shallow absorption line, not suited for radial-velocity measurements.  However, narrow absorption lines in similar positions in the \ion{Ca}{ii} H\&K line wings can be expected to similarly sample the upper photosphere, with dampened or even inverted granulation contrast.  Fig.\ \ref{fig:feicaii} shows the behavior of two such weak \ion{Fe}{i} absorption lines positioned in the flanks of, respectively, \ion{Ca}{ii}\,K and \ion{Ca}{ii}\,H.  These were selected as being clean and apparently unblended (except for their host lines), with their appearances in spectral atlases for both disk center and full-disk spectra shown at left.  Their behavior is different from ordinary lines in that the line cores are remarkably symmetric in both the non-magnetic and magnetic versions, seen as vertical bisector sections (although the upper bisector slopes turn sharply toward the red due to the sloping \ion{Ca}{ii} pseudocontinua).  The lack of any significant line-bottom asymmetry confirms that these lines are not formed in the deeper photosphere, but on levels where the granular velocity field has substantially lower amplitudes and/or less pronounced velocity-brightness correlations, and possibly also less variability.   

\section{Non-magnetic and magnetic radial velocities}

Fig.\ \ref{fig:lineshiftsall} summarizes the velocity shifts for all ordinary lines (i.e., excluding the very strong ones and those in the \ion{Ca}{ii} H\&K wings), as function of line depth and width.  The velocities are as obtained from the five-parameter fits illustrated in Fig.\ \ref{fig:linedepths}.  Convective velocities are dampened in the magnetic model, decreasing the velocity broadening and resulting in clearly narrower line profiles.  Among lines from the non-magnetic model, their convective blueshifts tend to be greater for weaker ones, but there is no striking variation with line-strength.  Differences between weak and strong lines are more manifest in their different asymmetries and bisector shapes.  Table \ref{table:1} gives detailed numbers while its averages over different wavelength regions shows some tendency for the shifts to decrease toward longer wavelengths, as expected from lower granulation contrast in the red.  The sample is too small to isolate higher-order dependences on excitation potential, as can otherwise be identified for idealized and isolated synthetic lines \citep{dravinsludwig23}. The unexpected and striking signature is the appearance of absolute redshifts in the magnetic spectra.  Their amounts are smaller than the convective blueshifts in nonmagnetic granulation (full-disk spectra averages -308 vs.\ +130 m\,s$^{-1}$) and -- similar to those -- are most pronounced at disk center, becoming more smeared out in full-disk spectra.

In the non-magnetic model, line-widths in full-disk spectra are broader than at disk center, reflecting the contributions from developed horizontal velocities which enhance Doppler broadening toward the limb, while the opposite is suggested in the magnetic model, apparently reflecting smaller amplitudes in horizontal velocity fields.      

 \begin{figure*}
\centering
 \includegraphics[width=18cm]{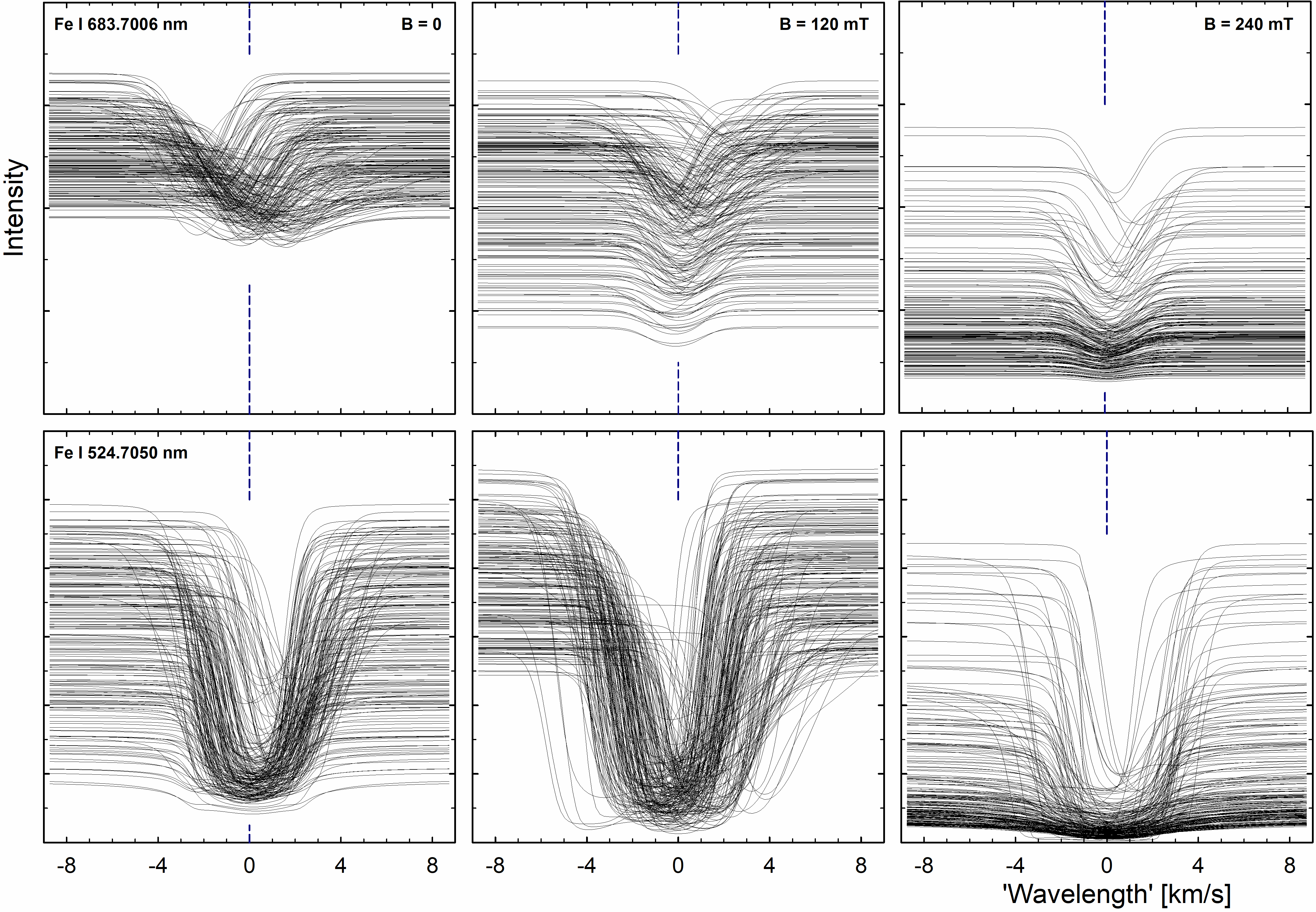}
     \caption{Spatially resolved line profiles across the simulation area at solar disk center illustrate how the convective blueshift in the field-free case gradually changes into a magnetic redshift.  Left to right, the weak \ion{Fe}{i} 683.7006 nm and strong \ion{Fe}{i} 524.7050 nm lines are shown (at the same pixels in the same snapshot) from simulations with fields = 0, 120, and 240 mT (0, 1200, 2400~G).  The nonmagnetic convective blueshift is caused by large, bright and blueshifted granules dominating over spatially narrow, dark and redshifted downflows.  In the magnetic case, surface areas of up- and downflows are more equal, velocities are smaller, while redshifted contributions from small bright elements bias the average. }  
\label{fig:spaghetti}
\end{figure*}
 
\section{Origin of redshifts in magnetic regions}

The line profiles examined so far result from extensive averaging over the simulation sequence and do not directly reveal the specific origin of the redshifts.  As studied in non-magnetic granulation, spatially resolved line profiles display highly varying shapes, reflecting their formation in an inhomogeneous medium \citep[e.g.,][]{asplund05, bergemannetal19, dravinsnordlund90}, while their averaged profiles and bisectors somehow encode their statistical distributions.  Tracing the origin of the magnetic redshifts requires a similarly detailed examination of the build-up of the averaged lines, where many thousands of spatially and temporally resolved profiles are combined.  

One representative temporal snapshot was selected and corresponding solar surface maps in Fig.\ \ref{fig:3dmodel} illustrate the lineshift variation across the model surface and its relation to various surface features.  In particular, contours of vertical velocity {\it{v}}$_z$ = 0 indicate locations where upflows turn into downflows.  A large fraction of the area is dominated by slow motions, where the convective velocities have been dampened by the magnetic fields.  Dynamically relevant features producing local redshifts are mostly associated with small areas of high intensity (`bright points'), so that averaged over space and time, the redshift injected from these localities causes a bias.  The dark inner parts of these bright points are associated with upflows (e.g., the feature at {\it{x,y}} = [3.7, 0.3] Mm), leading to a local blueshift of the line.  Remarkably, highest continuum brightness often coincides with the {\it{v}}$_z$ = 0 contour, indicating that heating (by adiabatic compression, shocks, friction) is occurring when gas is forced to turn over from narrow upflows into magnetically channeled downflows.  The width of these features is very small -- less than 100 km, and close to the resolution limit of these simulations, and require higher-resolution models to reveal further details. 

Spatially resolved profiles were calculated for 40\,$\times$\,40\, km$^{2}$ pixels, shown in Fig.\ \ref{fig:spaghetti} for the weak \ion{Fe}{i} 683.7006 nm and the strong line \ion{Fe}{i} 524.7050 nm line.  To limit cluttering, each plot shows only 1\% of the profiles in one snapshot, spatially uniformly sampled across the simulation area. To examine how the magnetic redshift signature evolves, line syntheses were carried out for models with successively higher magnetic fields. Besides for the currently examined 240 mT simulation, Fig.\ \ref{fig:spaghetti} also shows profiles from a corresponding field-free simulation, and one with a weaker 120 mT field.  The redshift signature develops gradually and the main mechanism is that magnetically channeled downflows become brighter in magnetoconvection -- even brighter than upflows.  Another effect that differs from the non-magnetic case, is that the surface area fractions of up- and downflows become more equal, or that downflow areas even dominate.  On some level of approximation, most of the surface can be seen as behaving rather neutral but the small bright points (despite subtending only a tiny fraction of the area) inject sufficiently many redshifted line components to bias the global average.  The origin of magnetic redshifts thus is qualitatively different from non-magnetic blueshifts, where the upflows are large, bright and blueshifted, but the downflows small, dark and redshifted, which results in a net blueshift of a line, with its familiar `C'-shaped bisector.

\section{Conclusions and outlook}

In this paper, line profiles and wavelength shifts arising in strongly magnetic granulation were examined.  Gradually varying solar surface area coverage by magnetic and nonmagnetic components enables to reproduce the general solar-cycle changes in its apparent radial velocity \citep{lakelandetal24, meunieretal10a}.  With detailed line profile and lineshift information, such variability can be more precisely modeled even if data for a mixture of magnetic field strengths may be needed.  Paper~I theoretically examined short-term radial-velocity jittering in non-magnetic granulation.  Forthcoming work will explore such short-term variability in magnetic granulation, with the aim to identify further measurable photospheric parameters that eventually could lead toward exoEarth detections. 

Finding exoEarths will be challenging but spectral-line wavelengths remain the most precisely measurable parameter, rather than line depth, width, or shape.  Optimal detection algorithms will profit from including the additional information contained in the differential behavior of individual spectral lines, as already experienced in observational studies \citep{almoullaetal22, almoullaetal24, annajohn25, cretignieretal20, demingetal24, dumusque18, meunieretal17, miklosetal20, thompsonetal20, wiseetal22}.  To fully understand how to optimally mitigate stellar radial-velocity fluctuations toward exoEarth detection, thus requires to understand the origins of specific line-profile variability and how their variations relate to wavelength displacements. 

Although simulations of 3D atmospheres now begin to reach a certain maturity, quite different levels of sophistication are possible for the ensuing computation of synthetic spectra, eventually perhaps envisioning `full' non-LTE treatments with Zeeman-sensitive line components for those lines with a non-zero Landé $\varg$$_{\textrm{eff}}$-factor, a connection to the lower chromosphere, and other.  The significance of non-LTE effects for also seemingly ordinary photospheric lines is suggested by studies of how different assumed levels of chromospherically influenced temperatures in the upper photosphere affect 1D non-LTE line formation in modeled G2~V stars \citep{vieytesetal25}.  The response to different levels of chromospheric heating is found to be markedly different among various \ion{Fe}{i} lines, even those with similar strengths and at nearby wavelengths, carrying a promise that those contain additional signatures of solar atmospheric modulation.  Such lines could be selected to observationally search for their variability although challenging non-LTE 3D calculations might be needed for their fuller understanding \citep{bergemannhoppe26, holzreuteretal25, lindamarsi24}.  Searches for planets around magnetically active stars face yet additional complications \citep[e.g.,][]{hebrardetal14}. 

The new PoET telescope at ESO VLT on Paranal is a significant addition to Sun-as-as-star facilities \citep{santosetal25}.  Besides providing spectra of the Sun-as-a-star with the ESPRESSO spectrometer's high resolving power approaching $\sim$200,000, it will be able to select limited portions of the solar disk to examine local line shapes and shifts, potentially enabling novel tests of magnetohydrodynamic model atmospheres.  The large ANDES spectrometer is being built for the ESO ELT \citep{palleetal25}.  While it will have outstanding light-collecting power, the challenges of interfacing a realistically large cross-dispersed échelle spectrometer to an extremely large telescope, limits its spectral resolution to similar ranges as current instruments.  In the longer term, one could certainly wish for also hyper-high resolutions of $\lambda$/$\Delta\lambda$ $\sim$1,000,000, corresponding to those in the present theoretical work, but then possibly requiring diffraction-limited adaptive-optics photonic instruments to limit their physical size.  The NIRPS radial-velocity spectrometer takes advantage of adaptive optics \citep{bouchyetal25} to limit its size but to realize the same at an extremely large telescope for the shorter visual wavelengths remains an interesting challenge.

\begin{acknowledgements}
{The work by DD is supported by grants from The Royal Physiographic Society of Lund.  CAP is thankful for funding from the Spanish government through grants AYA2014-56359-P, AYA2017-86389-P and PID2020-117493GB-100.  Important contributions to the development of the CO\,$^5$BOLD model family have been made by Bernd Freytag of Uppsala University.  Parts of this paper were completed by DD during a stay as a Scientific Visitor at the European Southern Observatory in Santiago de Chile.  Extensive use was made of NASA’s ADS Bibliographic Services and the arXiv$^{\circledR}$ distribution service.  We thank the anonymous referee for multiple insightful comments, which have improved the clarity of the text.  }

\end{acknowledgements}

\begin{appendix}

\nolinenumbers 

\section{Selected spectral lines}

Table \ref{table:1} lists the selected spectral lines, with wavelength values in air, excitation potentials $\chi$ in electron volts, and their Landé {$\varg$$_{\textrm{eff}}$}-factors; mostly from \citet{malherbe24}. Absolute line shifts at solar disk center, ${\mu}$=1, and for the integrated full disk are listed for non-magnetic and magnetic (240 mT) models.  The values are from five-parameter Gaussian-type fits, as described in the text and illustrated in Fig.\ \ref{fig:linedepths}.  Besides ordinary photospheric lines, the list also includes very strong lines from the green \ion{Mg}{i} triplet, the \ion{Na}{i} D lines, and the infrared \ion{Ca}{ii} triplet, as well as lines superposed onto the \ion{Ca}{ii}\,H \& K wings. Their wavelength shifts differ from those of ordinary lines and also differ between their cores and flanks.     

\end{appendix}

\begin{table*}
\tiny
\caption{Selected spectral lines, listing absolute line shifts at solar disk center and for the integrated full disk }     
        
\label{table:1}   
   
\centering          
\begin{tabular}{c c c c c c c c}
\hline  
Line [nm] & $\chi$ [eV] & Landé  & Non-magnetic  & Non-magnetic &Magnetic & Magnetic &\\ 
 &  &  $\varg$$_{\textrm{eff}}$  & ${\mu}$=1 [m\,s$^{-1}$]  & full disk [m\,s$^{-1}$] & ${\mu}$=1 [m\,s$^{-1}$] & full disk [m\,s$^{-1}$] & \\ 
\hline  
\\

\ion{Fe}{i} 389.2308   &  3.55 & ... &  -231 &  -192  & 257 &  211 &  \\ 
\ion{Fe}{i} 393.2627  &  2.73  & 1.1813 & 119  & 458  & 375  & 210 & Weak line in sloping Ca K line wing \\        
\ion{Fe}{i} 396.7421  &  3.30  &  0.826 & 116 & 348  & 394  & 211 & Weak line in sloping Ca H line wing \\        
\ion{Mn}{i} 426.5923  & 2.94  & 1.465 &  -428 & -374  & 397 & 160 \\        
\ion{Ca}{i} 431.8652  &  4.77  & 1.581 &  -407 & -306  & 402 & 126 & Very broad line\\       
\\ 
\ion{Fe}{ii} 449.1405  &  2.86  & 0.421 & -485 & -360  & 379 &160 & Strong line \\     
\ion{Fe}{ii} 450.8288  &  2.86  & 0.503 & -445 & -298 & 146 & -38 & Strong line \\        
\ion{Mg}{i} 457.1096  &  0.00  & 1.500 & -246 & -231 & 506 & 216 & Strong line \\ 
\ion{Fe}{ii} 465.6981  &  2.89  &  1.673 & -625 & -530 &  395& 137 \\ 
\ion{Fe}{ii} 466.6758   & 2.83  &  1.513 & -508 & -385 & 349 & 112 \\ 
\\
\ion{Fe}{ii} 467.0182  &  2.58  & 1.169 &  -545 & -426  & 436 & 182 \\ 
\ion{Mg}{i} 517.2684   & 2.71  & 1.750 & 148 & 358  & 354 & 193 & Green Mg triplet  \ion{Mg}{i} b$_2$ \\ 
\ion{Mg}{i} 518.3604  &  2.72  & 1.250 & 183 & 403  &  356 & 196 & Green Mg triplet \ion{Mg}{i} b$_1$ \\ 
\ion{Fe}{i} 524.2491  &  3.63  & 1.004 & -393 & -327  & 303 &115  & Strong line \\ 
\ion{Cr}{ii} 524.6768  & 3.71  & ... & -616  & -498  & 413 & 127  & Very weak line \\   
\\    
\ion{Fe}{i} 524.7050  &  0.09  & 1.992 & -244&  -226  & 393 & 172  \\	
\ion{Ti}{i} 524.7289  & 4.47  & ... & -439  &-366 & 262 & 117 \\
\ion{Cr}{i} 524.7565   & 0.96  & 2.512 & -269 & -228 & 429 & 187  \\
\ion{Co}{i} 524.7911  & 4.15  & ... & -361 & -312 &  249 & 120 & Very weak line \\
\ion{Fe}{i} 525.0209  & 0.12  & 2.999 & -476 & -456  & 164 & -59  \\
\\
\ion{Fe}{i} 525.3021   & 2.28  & 1.008 & -418 & -355  & 236 & 93 & Very weak line   \\
\ion{Na}{i} 588.9951  & 0.00  & 1.167 & 66 & 271  & 376 & 201 & \ion{Na}{i} D$_2$ \\
\ion{Na}{i} 589.5924   & 0.00  & 1.333 & 43 & 218  &  380 & 202 & \ion{Na}{i} D$_1$ core \\
\ion{Na}{i} 589.5924   & 0.00  & 1.333 & -113 & 30  &  437 & 210 & \ion{Na}{i} D$_1$ full \\
\ion{Fe}{i} 672.5356  & 5.95  & 1.392 & -393  & -335 & 239 & 83 & Very weak line \\
\\
\ion{Ni}{i} 676.7768   & 1.83  & 1.425 & -247& -187  & 356 & 159  \\
\ion{Fe}{i} 683.7006  &  4.59 & 1.150 & -421& -353  & 282 & 122 & Very weak line  \\
\ion{Fe}{i} 741.8667  & 4.14  & 0.872 & -317 & -242  &  343 & 172  \\
\ion{Fe}{i} 751.1019  & 5.83  & 1.419 & -206  & -71  & 380 & 191  \\
\ion{Fe}{i} 754.0430  & 2.73  & ... & -295 & -215  & 269 & 148 & Very weak line \\
\\
\ion{Ni}{i} 755.5598  & 5.49  &  1.006 & -260 & -163  & 333 & 151  \\
\ion{Ca}{ii} 849.8023  & 1.69  & 1.067 & 223 & 507 & ...  & ...  & IR Ca triplet core \\
\ion{Ca}{ii} 849.8023  & 1.69  & 1.067 & 95 & 393 & 296  & 113  & IR Ca triplet flanks \\
\ion{Ca}{ii} 854.2091  & 1.70  & 1.099 & 612 & 955  & ...  & ... & IR Ca triplet core \\
\ion{Ca}{ii} 854.2091  & 1.70  & 1.099 & 280 & 537  & 83 & -45 & IR Ca triplet flanks\\
\\
\ion{Ti}{i} 873.4712  & 1.05  & ... & -483 & -434  &  152 & 75 & Very weak line  \\
\ion{Mg}{i} 873.6019  & 5.95  & 1.125 & -349  & -258 & 254 & 74  \\
\ion{Fe}{i} 944.3801  &  6.40  & ... &  -262 & -173  & 361  & 193  \\
\\
Average & 3.28  & ... & -384 & -308 & 321 &130 & Average over all ordinary lines \\
$\langle$\,350-500 nm\,$\rangle$  & 2.81 & ... & -435  & -345 & 363 & 141 & Average over ordinary lines \\
$\langle\,$500-700 nm$\,\rangle$ & 2.89 & ...  & -389 & -331 & 302 & 112 & Average over ordinary lines \\
$\langle$\,700-950 nm$\,\rangle$ & 4.51 & ... & -310 & -222 & 298 & 143 & Average over ordinary lines \\
\\
\hline       
\end{tabular}
\label{table:linelist1}
\end{table*}


\begin{thebibliography}{}

\bibitem[Al Moulla et al.(2022)]{almoullaetal22} Al Moulla, K., Dumusque, X., Cretignier, M., et al.\ 2022, \aap, 664, A34 
\bibitem[Al Moulla et al.(2024)]{almoullaetal24} Al Moulla, K., Dumusque, X., \& Cretignier, M.\ 2024, \aap, 683, A106 
\bibitem[Anna John et al.(2025)]{annajohn25} Anna John, A., Al Moulla, K., O'Sullivan, N.~K., et al.\ 2025, \mnras, 543, 1974 
\bibitem[Artigau et al.(2022)]{artigauetal22} Artigau, {\'E}., Cadieux, C., Cook, N.~J., et al.\ 2022, \aj, 164, 84 
\bibitem[Asplund(2005)]{asplund05} Asplund, M.\ 2005, \araa, 43, 481 
\bibitem[Beeck et al.(2015a)]{beecketal15a} Beeck, B., Sch{\"u}ssler, M., Cameron, R.~H., et al.\ 2015a, \aap, 581, A42 
\bibitem[Beeck et al.(2015b)]{beecketal15b} Beeck, B., Sch{\"u}ssler, M., Cameron, R.~H., et al.\ 2015b, \aap, 581, A43 
\bibitem[Bellot Rubio \& Orozco Su{\'a}rez(2019)]{bellotrubioorozcosuarez19} Bellot Rubio, L. \& Orozco Su{\'a}rez, D.\ 2019, Liv.\ Rev.\ Solar Phys., 16, 1 
\bibitem[Bergemann et al.(2019)]{bergemannetal19} Bergemann, M., Gallagher, A.~J., Eitner, P., et al.\ 2019, \aap, 631, A80 
\bibitem[Bergemann \& Hoppe(2026)]{bergemannhoppe26} Bergemann, M. \& Hoppe, R.\ 2026, Liv.\ Rev.\ Comp.\ Astroph., in press,\\ arXiv:2511.04254 
\bibitem[Berger et al.(2004)]{bergeretal04} Berger, T.~E., Rouppe van der Voort, L.~H.~M., L{\"o}fdahl, M.~G., et al.\ 2004, \aap, 428, 613 
\bibitem[Bhatia et al.(2022)]{bhatiaetal22} Bhatia, T.~S., Cameron, R.~H., Solanki, S.~K., et al.\ 2022, \aap, 663, A166  
\bibitem[Bhatia et al.(2026)]{bhatiaetal26} Bhatia, T.S., Cameron, R.~H., Solanki, S.~K., et al.\ 2026, \aap, 706, A308 
\bibitem[Blackman et al.(2020)]{blackmanetal20} Blackman, R.~T., Fischer, D.~A., Jurgenson, C.~A., et al.\ 2020, \aj, 159, 238 
\bibitem[Bouchy et al.(2025)]{bouchyetal25} Bouchy, F., Doyon, R., Pepe, F., et al.\ 2025, \aap, 700, A10 
\bibitem[Brandt \& Solanki(1990)]{brandtsolanki90} Brandt, P.~N. \& Solanki, S.~K.\ 1990, \aap, 231, 221  
\bibitem[Cavallini et al.(1985)]{cavallinietal85} Cavallini, F., Ceppatelli, G., \& Righini, A.\ 1985, \aap, 143, 116 
\bibitem[Cavallini et al.(1986)]{cavallinietal86} Cavallini, F., Ceppatelli, G., \& Righini, A.\ 1986, \aap, 158, 275 
\bibitem[Cavallini et al.(1988)]{cavallinietal88} Cavallini, F., Ceppatelli, G., \& Righini, A.\ 1988, \aap, 205, 278 
\bibitem[Cavallini et al.(1989)]{cavallinietal89} Cavallini, F., Ceppatelli, G., \& Righini, A.\ 1989, Solar and Stellar Granulation, NATO ASI 263, 283  
\bibitem[Cegla et al.(2012)]{ceglaetal12} Cegla, H.~M., Watson, C.~A., Marsh, T.~R., et al.\ 2012, \mnras, 421, L54 
\bibitem[Cegla et al.(2019)]{ceglaetal19} Cegla, H.~M., Watson, C.~A., Shelyag, S., et al.\ 2019, \apj, 879, 55 
\bibitem[Chiavassa et al.(2018)]{chiavassaetal18} Chiavassa, A., Casagrande, L., Collet, R., et al.\ 2018, \aap, 611, A11 
\bibitem[Crass et al.(2021)]{crassetal21} Crass, J., Gaudi, B.~S., Leifer, S., et al.\ 2021, Extreme Precision Radial Velocity Working Group Final Report, arXiv:2107.14291
\bibitem[Cretignier et al.(2020)]{cretignieretal20} Cretignier, M., Dumusque, X., Allart, R., et al.\ 2020, \aap, 633, A76 
\bibitem[Cretignier et al.(2021)]{cretignieretal21} Cretignier, M., Dumusque, X., Hara, N.~C., et al.\ 2021, \aap, 653, A43 
\bibitem[Delbouille et al. (1989)]{delbouilleetal89} Delbouille, L., Roland, G., \& Neven, L.\ 1989, Atlas photom\'{e}trique du spectre solaire de $\lambda$~3000 a $\lambda$~10000 (Li\`{e}ge: Universit\'{e} de Li\`{e}ge, Institut d'Astrophysique) [`Jungfraujoch atlas'] \\ Digital version:  \url{http://bass2000.obspm.fr}
\bibitem[Deming et al.(2024)]{demingetal24} Deming, D., Llama, J., \& Fu, G.\ 2024, \aj, 167, 34  
\bibitem[de Wijn et al.(2009)]{dewijnetal09} de Wijn, A.~G., Stenflo, J.~O., Solanki, S.~K., et al.\ 2009, \ssr, 144, 275 
\bibitem[Doyon et al.(2025)]{doyonetal25} Doyon, R., Bouchy, F., Pepe, F., et al.\ 2025, The Messenger, 194, 13 
\bibitem[Dravins \& Ludwig(2023)]{dravinsludwig23} Dravins, D., Ludwig, H.-G.\ 2023, \aap, 679, A3 (Paper I)
\bibitem[Dravins \& Ludwig(2024)]{dravinsludwig24} Dravins, D., Ludwig, H.-G.\ 2024, \aap, 687, A60 (Paper II)
\bibitem[Dravins \& Nordlund(1990)]{dravinsnordlund90} Dravins, D., \& Nordlund, {\AA}.\ 1990, \aap, 228, 184 
\bibitem[Dravins et al.(2021a)]{dravinsetal21a} Dravins, D., Ludwig, H.-G., \& Freytag, B.\ 2021a, \aap, 649, A16 
\bibitem[Dravins et al.(2021b)]{dravinsetal21b} Dravins, D., Ludwig, H.-G., \& Freytag, B.\ 2021b, \aap, 649, A17 
\bibitem[Dumusque(2018)]{dumusque18} Dumusque, X.\ 2018, \aap, 620, A47 
\bibitem[Fischer et al.(2016)]{fischeretal16} Fischer, D.~A., Anglada-Escude, G., Arriagada, P., et al.\ 2016, \pasp, 128, 066001 
\bibitem[Ford et al.(2024)]{fordetal24} Ford, E.~B., Bender, C.~F., Blake, C.~H., et al.\ 2024, submitted to AAS journals, arXiv:2408.13318 
\bibitem[Frame et al.(2025)]{frameetal25} Frame, G., Cegla, H.~M., Witzke, V., et al.\ 2025, \mnras, 539, 2248 
\bibitem[Freytag et al.(2012)]{freytagetal12} Freytag, B., Steffen, M., Ludwig, H.-G., et al.\ 2012, J.\ Comp.\ Phys., 231, 919 
\bibitem[Gupta \& Bedell(2024)]{guptabedell24} Gupta, A.~F. \& Bedell, M.\ 2024, \aj, 168, 29. 
\bibitem[Hahlin et al.(2023)]{hahlinetal23} Hahlin, A., Kochukhov, O., Rains, A.~D., et al.\ 2023, \aap, 675, A91 
\bibitem[Hall et al.(2018)]{halletal18} Hall, R.~D., Thompson, S.~J., Handley, W., et al.\ 2018, \mnras, 479, 2968 
\bibitem[H{\'e}brard et al.(2014)]{hebrardetal14} H{\'e}brard, {\'E}. M., Donati, J.-F., Delfosse, X., et al.\ 2014, \mnras, 443, 2599 
\bibitem[Heiter et al.(2015)]{heiteretal15} Heiter, U., Lind, K., Asplund, M., et al.\ 2015, \physscr, 90, 054010 
\bibitem[Hinkle et al.(2000)]{hinkleetal00} Hinkle, K., Wallace, L., Valenti, J., Harmer, D.\ 2000, Visible and Near Infrared Atlas of the Arcturus Spectrum 3727-9300 Å, Astron.\ Soc.\ Pacific, \\ San Francisco
\bibitem[Holzreuter \& Solanki(2012)]{holzreutersolanki12} Holzreuter, R. \& Solanki, S.~K.\ 2012, \aap, 547, A4
\bibitem[Holzreuter \& Solanki(2015)]{holzreutersolanki15} Holzreuter, R. \& Solanki, S.~K.\ 2015, \aap, 582, A101 
\bibitem[Holzreuter et al.(2025)]{holzreuteretal25} Holzreuter, R., Smitha, H.~N., \& Solanki, S.~K.\ 2025, \aap, 697, A105 
\bibitem[Immerschitt \& Schroeter(1989)]{immerschitt_schroter89} Immerschitt, S. \& Schroeter, E.~H.\ 1989, \aap, 208, 307  
\bibitem[Ishikawa et al.(2007)]{ishikawaetal07} Ishikawa, R., Tsuneta, S., Kitakoshi, Y., et al.\ 2007, \aap, 472, 911 
\bibitem[Jess et al.(2010)]{jessetal10} Jess, D.~B., Mathioudakis, M., Christian, D.~J., et al.\ 2010, \apjl, 719, L134 
\bibitem[Keys et al.(2013)]{keysetal13} Keys, P.~H., Mathioudakis, M., Jess, D.~B., et al.\ 2013, \mnras, 428, 3220 
\bibitem[Keys et al.(2026)]{keysetal26} Keys, P.~H., Campbell, R.~J., Magill, D.~K.~J., et al.\ 2026, \apj, 999, 201 
\bibitem[Khomenko et al.(2018)]{khomenkoetal18} Khomenko, E., Vitas, N., Collados, M., et al.\ 2018, \aap, 618, A87 
\bibitem[Koesterke et al.(2008)]{koesterkeetal08} Koesterke, L., Allende Prieto, C., \& Lambert, D.~L.\ 2008, \apj, 680, 764 
\bibitem[Komori et al.(2025)]{komorietal25} Komori, C., Brewer, J.~M., \& Zhao, L.~L.\ 2025, \aj, 170, 209 
\bibitem[Kostik \& Khomenko(2012)]{kostikkhomenko12} Kostik, R. \& Khomenko, E.~V.\ 2012, \aap, 545, A22 
\bibitem[Krikova et al.(2023)]{krikovaetal23} Krikova, K., Pereira, T.~M.~D., \& Rouppe van der Voort, L.~H.~M.\ 2023, \aap, 677, A52  
\bibitem[Kuridze et al.(2025)]{kuridzeetal25} Kuridze, D., W{\"o}ger, F., Uitenbroek, H., et al.\ 2025, \apjl, 985, L23 
\bibitem[Kurucz et al.(1984)]{kuruczetal84} Kurucz, R.~L., Furenlid, I., Brault, J., Testerman, L.\ 1984, National Solar Observatory Atlas, Sunspot, New Mexico: National Solar Observatory [`Kitt Peak atlas']  Digital version: \url{http://diglib.nso.edu} 
\bibitem[Lagg et al.(2016)]{laggetal16} Lagg, A., Solanki, S.~K., Doerr, H.-P., et al.\ 2016, \aap, 596, A6 
\bibitem[Lakeland et al.(2024)]{lakelandetal24} Lakeland, B.~S., Naylor, T., Haywood, R.~D., et al.\ 2024, \mnras, 527, 7681 
\bibitem[Lind \& Amarsi(2024)]{lindamarsi24} Lind, K. \& Amarsi, A.~M.\ 2024, \araa, 62, 475 
\bibitem[Lindegren \& Dravins(2003)]{lindegrendravins03} Lindegren, L. \& Dravins, D.\ 2003, \aap, 401, 1185 
\bibitem[Lo Curto et al.(2024)]{locurtoetal24} Lo Curto, G., Pepe, F., Fleury, M., et al.\ 2024, The Messenger, 192, 38  
\bibitem[Ludwig et al.(2023)]{ludwigetal23} Ludwig, H.-G., Steffen, M., \& Freytag, B.\ 2023, \aap, 679, A65 
\bibitem[Malherbe(2024)]{malherbe24} Malherbe, J.-M.\ 2024, A compilation of solar atlases, arXiv:2404.16902  
\bibitem[Mayor et al.(2003)]{mayoretal03} Mayor, M., Pepe, F., Queloz, D., et al.\ 2003, The Messenger, 114, 20. 
\bibitem[Meunier \& Delfosse(2009)]{meunierdelfosse09} Meunier, N. \& Delfosse, X.\ 2009, \aap, 501, 1103 
\bibitem[Meunier \& Sulis(2026)]{meuniersulis26} Meunier, N. \& Sulis, S.\ 2026, \aap, 707, A187 
\bibitem[Meunier et al.(2010a)]{meunieretal10a} Meunier, N., Desort, M., \& Lagrange, A.-M.\ 2010a, \aap, 512, A39 
\bibitem[Meunier et al.(2010b)]{meunieretal10b} Meunier, N., Lagrange, A.-M., \& Desort, M.\ 2010b, \aap, 519, A66 
\bibitem[Meunier et al.(2017)]{meunieretal17} Meunier, N., Lagrange, A.-M., \& Borgniet, S.\ 2017, \aap, 607, A6 
\bibitem[Meunier et al.(2024)]{meunieretal24} Meunier, N., Lagrange, A.-M., Dumusque, X., et al.\ 2024, \aap, 687, A303 
\bibitem[Miklos et al.(2020)]{miklosetal20} Miklos, M., Milbourne, T.~W., Haywood, R.~D., et al.\ 2020, \apj, 888, 117 
\bibitem[Norris et al.(2023)]{norrisetal23} Norris, C.~M., Unruh, Y.~C., Witzke, V., et al.\ 2023, \mnras, 524, 1139 
\bibitem[O'Sullivan \& Aigrain(2024)]{osullivanaigrain24} O'Sullivan, N.~K. \& Aigrain, S.\ 2024, \mnras, 531, 4181 
\bibitem[Palle et al.(2025)]{palleetal25} Palle, E., Biazzo, K., Bolmont, E., et al.\ 2025, Exp.\ Astron., 59, 29 
\bibitem[Palumbo et al.(2022)]{palumboetal22} Palumbo, M.~L., Ford, E.~B., Wright, J.~T., et al.\ 2022, \aj, 163, 11 
\bibitem[Panja et al.(2020)]{panjaetal20} Panja, M., Cameron, R., \& Solanki, S.~K.\ 2020, \apj, 893, 113 
\bibitem[Rackham et al.(2023)]{rackhametal23} Rackham, B.~V., Espinoza, N., Berdyugina, S.~V., et al.\ 2023, RAS Techn.\ \\ Instrum., 2, 148 
\bibitem[Reiners et al.(2016)]{reinersetal16} Reiners, A., Mrotzek, N., Lemke, U., et al.\ 2016, \aap, 587, A65 [`Göttingen atlas'] 
\bibitem[Romano et al.(2012)]{romanoetal12} Romano, P., Berrilli, F., Criscuoli, S., et al.\ 2012, \solphys, 280, 407 
\bibitem[Rouppe van der Voort et al.(2005)]{rouppevandervoortetal05} Rouppe van der Voort, L.~H.~M., Hansteen, V.~H., Carlsson, M., et al.\ 2005, \aap, 435, 327 
\bibitem[Ryabchikova et al.(2015)]{ryabchikovaetal15} Ryabchikova, T., Piskunov, N., Kurucz, R.~L., et al.\ 2015, \physscr, 90, 054005 
\bibitem[Quintero Noda et al.(2021)]{quinteronodaetal21} Quintero Noda, C., Barklem, P.~S., Gafeira, R., et al.\ 2021, \aap, 652, A161 
\bibitem[Salhab et al.(2018)]{salhabetal18} Salhab, R.~G., Steiner, O., Berdyugina, S.~V., et al.\ 2018, \aap, 614, A78 
\bibitem[Salzer et al.(2025)]{salzeretal25} Salzer, J., Cisewski-Kehe, J., Ford, E.~B., et al.\ 2025, \aj, 170, 179 
\bibitem[Santos et al.(2025)]{santosetal25} Santos, N.~C., Cabral, A., Leite, I., et al.\ 2025, The Messenger, 194, 21
\bibitem[Sch{\"u}ssler \& V{\"o}gler(2006)]{schusslervogler06} Sch{\"u}ssler, M. \& V{\"o}gler, A.\ 2006, \apjl, 641, L73 
\bibitem[Sheminova(2022)]{sheminova22} Sheminova, V.~A.\ 2022, Kin.\ Phys.\ Celestial Bodies, 38, 83 
\bibitem[Shukla et al.(2026)]{shuklaetal26} Shukla, M., Pandit, S., \& Yadav, N.\ 2026, \mnras, 547, 1  
\bibitem[Sinjan et al.(2024)]{sinjanetal24} Sinjan, J., Solanki, S.~K., Hirzberger, J., et al.\ 2024, \aap, 690, A341  
\bibitem[Smitha \& Solanki(2017)]{smithasolanki17} Smitha, H.~N. \& Solanki, S.~K.\ 2017, \aap, 608, A111   
\bibitem[Smitha et al.(2021)]{smithaetal21} Smitha, H.~N., Holzreuter, R., van Noort, M., et al.\ 2021, \aap, 647, A46  
\bibitem[Smitha et al.(2025)]{smithaetal25} Smitha, H.~N., Shapiro, A.~I., Witzke, V., et al.\ 2025, \apjl, 978, L13 
\bibitem[Solanki(1993)]{solanki93} Solanki, S.~K.\ 1993, \ssr, 63, 1 
\bibitem[Stein(2012)]{stein12} Stein, R.~F.\ 2012, Living Rev.\ Solar Phys., 9, 4 
\bibitem[Uitenbroek(2006)]{uitenbroek06} Uitenbroek, H.\ 2006, \apj, 639, 516  
\bibitem[Uitenbroek \& Tritschler(2006)]{uitenbroektrischler06} Uitenbroek, H. \& Tritschler, A.\ 2006, \apj, 639, 525  
\bibitem[Thompson et al.(2020)]{thompsonetal20} Thompson, A.~P.~G., Watson, C.~A., Haywood, R.~D., et al.\ 2020, \mnras, 494, 4279 
\bibitem[Tremblay et al.(2013)]{tremblayetal13} Tremblay, P.-E., Ludwig, H.-G., Freytag, B., et al.\ 2013, \aap, 557, A7 
\bibitem[Tritschler \& Uitenbroek(2006)]{tritschleruitenbroek06} Tritschler, A. \& Uitenbroek, H.\ 2006, \apj, 648, 741 
\bibitem[Vieytes et al.(2025)]{vieytesetal25} Vieytes, M.~C., Zhao, L.~L., \& Bedell, M.\ 2025, \apj, 981, 4 
\bibitem[V{\"o}gler et al.(2005)]{vogleretal05} V{\"o}gler, A., Shelyag, S., Sch{\"u}ssler, M., et al.\ 2005, \aap, 429, 335  
\bibitem[Wilken et al.(2012)]{wilkenetal12} Wilken, T., Curto, G.~L., Probst, R.~A., et al.\ 2012, \nat, 485, 611
\bibitem[Wilson(1938)]{wilson38} Wilson, O.~C.\ 1938, \pasp, 50, 245 
\bibitem[Wise et al.(2022)]{wiseetal22} Wise, A., Plavchan, P., Dumusque, X., et al.\ 2022, \apj, 930, 121 
\bibitem[Zhao et al.(2026)]{zhaoetal26} Zhao, L.~L., Fischer, D.~A., Szymkowiak, A.~E., et al.\ 2026, \apjs, 282, 71  

\end{thebibliography}
\end{document}